\documentclass[12pt,a4paper]{article}
\usepackage{amssymb}
\usepackage[dvips]{graphicx} 
\begin{document}
\begin{titlepage}
\begin{flushright}
UAB--FT--534\\
hep-ph/0606314\\
November 2006
\end{flushright}
\vspace*{1.6cm}

\begin{center}
{\Large\bf 
Scalar and vector meson exchange in $V\rightarrow P^0P^0\gamma$ decays}\\
\vspace*{0.8cm}

R.~Escribano$^1$\\
\vspace*{0.2cm}

{\footnotesize\it
$^1$Grup de F\'{\i}sica Te\`orica and IFAE,
Universitat Aut\`onoma de Barcelona,\\
E-08193 Bellaterra (Barcelona), Spain}

\end{center}
\vspace*{0.2cm}

\begin{abstract}
    The scalar contributions to the radiative decays of light vector mesons into
    a pair of neutral pseudoscalars, $V\rightarrow P^0P^0\gamma$,
    are studied within the framework of the Linear Sigma Model.
    This model has the advantage of incorporating not only the scalar resonances in
    an explicit way but also the constraints required by chiral symmetry.
    The experimental data on $\phi\rightarrow\pi^0\pi^0\gamma$,
    $\phi\rightarrow\pi^0\eta\gamma$, $\rho\rightarrow\pi^0\pi^0\gamma$ and
    $\omega\rightarrow\pi^0\pi^0\gamma$ are satisfactorily accommodated in our framework.
    Theoretical predictions for $\phi\rightarrow K^0\bar K^0\gamma$,
    $\rho\rightarrow\pi^0\eta\gamma$, $\omega\rightarrow\pi^0\eta\gamma$
    and the ratio $\phi\to f_0\gamma/a_0\gamma$ are also given.
\end{abstract}
\end{titlepage}

\section{Introduction}
\label{intro}
The radiative decays of the light vector mesons $(V=\rho,\omega,\phi)$
into a pair of neutral pseudoscalars $(P=\pi^0,K^0,\eta)$,
$V\rightarrow P^0P^0\gamma$, are an excellent laboratory for investigating the
nature and extracting the properties of the light scalar meson resonances
$(S=\sigma,a_{0},f_{0})$.
The reason is the following: looking at the quantum numbers of the initial vector
and those of the final photon, both with $J^{PC}=1^{--}$, the system made of the
two neutral pseudoscalars are mainly in a $0^{++}$ state, {\it i.e.}~a scalar state,
or $2^{++}$. However, the lightest tensor resonances have masses of the order of
1.2 GeV and therefore their contributions to these processes are supposed to be
negligible.
In addition, there is also a vector meson contribution when one of the neutral
pseudoscalars and the photon are produced by the exchange of an intermediate
vector meson through the decay chain $V\rightarrow VP^0\rightarrow P^0P^0\gamma$.
Fortunately, for most of the processes of interest the main contribution is by 
far the scalar one, thus making of the study of these radiative decays a very
challenging subject in order to improve our knowledge on the lightest scalar 
mesons.
This study also complements other analysis based on central production,
$D$ and $J/\psi$ decays, etc.~\cite{Close:2002zu}.
Particularly interesting are the so called \emph{golden processes}, namely
$\phi\rightarrow\pi^0\pi^0\gamma$, $\phi\rightarrow\pi^0\eta\gamma$ and
$\rho\rightarrow\pi^0\pi^0\gamma$, which, as we will see, can provide us with
valuable information on the properties of the $f_{0}(980)$, $a_{0}(980)$ and
$\sigma(600)$ resonances, respectively.
A further motivation for the present work is the renewed interest on these
radiative decays from both the theoretical
\cite{Isidori:2006we,Black:2006mn,Achasov:2005hm}
and experimental \cite{Ambrosino:2006gk,Eidelman:2006} sides.

At present, there are two experimental facilities that provide measurements on
the $V\rightarrow P^0P^0\gamma$ decays.
One is the VEPP-2M $e^+e^-$ collider in Novosibirsk with two experimental
groups SND and CMD-2, and the other is the DA$\Phi$NE $\phi$-factory in Frascati
with the KLOE Collaboration.
The Russian experiment operates at different center of mass energies having the
advantage of not only measuring the processes $\phi\rightarrow\pi^0\pi^0\gamma$
and $\phi\rightarrow\pi^0\eta\gamma$ but also $\rho\rightarrow\pi^0\pi^0\gamma$
and other similar decays.
On the contrary, the Italian experiment operates at a fixed center of mass energy
around the $\phi$ mass and is only able, at least in principle, to measure the
$\phi$-decay processes but not others.
However, due to its higher luminosity the statistical accuracy of DA$\Phi$NE
measurements is better than in VEPP-2M.
Moreover, the good performance of the KLOE detector makes also the systematic
error smaller and as a consequence the DA$\Phi$NE measurements on $\phi$-decay
processes are in general more precise.
For $\phi\rightarrow\pi^0\pi^0\gamma$, the first measurements of this decay
have been reported by the SND and CMD-2 Collaborations.
For the branching ratio they obtain
$B(\phi\rightarrow\pi^0\pi^0\gamma)=(1.221\pm 0.098\pm 0.061)\times 10^{-4}$
\cite{Achasov:2000ym} and $(0.92\pm 0.08\pm 0.06)\times 10^{-4}$ \cite{Akhmetshin:1999di},
for $m_{\pi\pi}>700$ MeV in the latter case.
More recently, the KLOE Collaboration has measured
$B(\phi\rightarrow\pi^0\pi^0\gamma)=(1.09\pm 0.03\pm 0.05)\times 10^{-4}$
\cite{Aloisio:2002bt} in agreement with VEPP-2M results but with a considerably
smaller error.
In all the cases, the spectrum is clearly peaked at $m_{\pi\pi}\simeq 970$ MeV,
as expected from an important $f_{0}(980)$ contribution.
For $\phi\rightarrow\pi^0\eta\gamma$, the branching ratios measured by the
SND and CMD-2 Collaborations are
$B(\phi\rightarrow\pi^0\eta\gamma)=(8.8\pm 1.4\pm 0.9)\times 10^{-5}$
\cite{Achasov:2000ku} and $(9.0\pm 2.4\pm 1.0)\times 10^{-5}$ \cite{Akhmetshin:1999di},
and by the KLOE Collaboration are
$B(\phi\rightarrow\pi^0\eta\gamma)=(8.51\pm 0.51\pm 0.57)\times 10^{-5}$
from $\eta\rightarrow\gamma\gamma$ and $(7.96\pm 0.60\pm 0.40)\times 10^{-5}$
from $\eta\rightarrow\pi^+\pi^-\pi^0$ \cite{Aloisio:2002bs}.
The two values are in agreement and also agree with those of VEPP-2M.
Again, in all the cases, the observed invariant mass distribution shows a
significant enhancement at large $\pi^0\eta$ invariant mass that is
interpreted as a manifestation of the dominant contribution of the
$a_{0}\gamma$ intermediate state.
For $\rho\rightarrow\pi^0\pi^0\gamma$, the only existing measurements in the
literature come from the study of the $e^+e^-\rightarrow\pi^0\pi^0\gamma$ process
by the SND and CMD-2 experiments in the energy region 0.60--0.97 GeV.
From the analysis of the energy dependence of the measured cross section
they obtain for the branching ratio
$B(\rho\rightarrow\pi^0\pi^0\gamma)=(4.1^{+1.0}_{-0.9}\pm 0.3)\times 10^{-5}$
\cite{Achasov:2002jv} and
$B(\rho\rightarrow\pi^0\pi^0\gamma)=(5.2^{+1.5}_{-1.3}\pm 0.6)\times 10^{-5}$
\cite{Akhmetshin:2003rg}, in agreement with the older result
$B(\rho\rightarrow\pi^0\pi^0\gamma)=(4.8^{+3.4}_{-1.8}\pm 0.2)\times 10^{-5}$
\cite{Achasov:2000zr}.
These values can be explained by means of a significant contribution of the
$\sigma\gamma$ intermediate state together with the well-known $\omega\pi$ contribution.
For $\omega\rightarrow\pi^0\pi^0\gamma$, the values of the branching ratio
$B(\omega\rightarrow\pi^0\pi^0\gamma)=(6.6^{+1.4}_{-1.3}\pm 0.6)\times 10^{-5}$
\cite{Achasov:2002jv} and
$B(\omega\rightarrow\pi^0\pi^0\gamma)=(6.4^{+2.4}_{-2.0}\pm 0.8)\times 10^{-5}$
\cite{Akhmetshin:2003rg}
are found in agreement with the older GAMS result
$B(\omega\rightarrow\pi^0\pi^0\gamma)=(7.2\pm 2.6)\times 10^{-5}$ \cite{Alde:1994kf}.
Finally, an upper limit for $\omega\rightarrow\pi^0\eta\gamma$ has been obtained:
$B(\omega\rightarrow\pi^0\eta\gamma)<3.3\times 10^{-5}$ at 90\% CL \cite{Akhmetshin:2003rg}.
For the rest of the processes, namely $\phi\rightarrow K^0\bar K^0\gamma$ and $\rho\rightarrow\pi^0\eta\gamma$,
the branching ratios are predicted to be very small and have not been yet measured.

An early attempt to explain the $V\rightarrow P^0P^0\gamma$ decays was done in
Ref.~\cite{Bramon:1992kr} using the vector meson dominance (VMD) model.
In this framework, the $V\rightarrow P^0P^0\gamma$ decays proceed through the
decay chain $V\rightarrow VP^0\rightarrow P^0P^0\gamma$, where the intermediate
vector mesons exchanged are
$V=\rho$ for $(\omega,\phi)\rightarrow\pi^0\pi^0\gamma$,
$V=\omega$ for $\rho\rightarrow\pi^0\pi^0\gamma$,
$V=K^{\ast 0}, \bar K^{\ast 0}$ for $\phi\rightarrow K^0\bar K^0\gamma$,
$V=\rho, \omega$
for $(\rho,\omega)\rightarrow\pi^0\eta\gamma$, and
$V=\rho, \omega$
for $\phi\rightarrow\pi^0\eta\gamma$.
The calculated branching ratios are
$B_{\phi\rightarrow\pi^0\pi^0\gamma}^{\rm VMD}=1.2\times 10^{-5}$,
$B_{\phi\rightarrow\pi^0\eta\gamma}^{\rm VMD}=5.4\times 10^{-6}$,
$B_{\rho\rightarrow\pi^0\pi^0\gamma}^{\rm VMD}=1.1\times 10^{-5}$,
$B_{\omega\rightarrow\pi^0\pi^0\gamma}^{\rm VMD}=2.8\times 10^{-5}$,
$B_{\phi\rightarrow K^0\bar K^0\gamma}^{\rm VMD}=2.7\times 10^{-12}$,
$B_{\rho\rightarrow\pi^0\eta\gamma}^{\rm VMD}=4\times 10^{-10}$ and
$B_{\omega\rightarrow\pi^0\eta\gamma}^{\rm VMD}=1.6\times 10^{-7}$
\cite{Bramon:1992kr}.
The first four are found to be substantially smaller than the experimental results
quoted before.
Later on, the same authors studied the $V\rightarrow P^0P^0\gamma$ decays in a 
Chiral Perturbation Theory (ChPT) context enlarged to include on-shell vector
mesons \cite{Bramon:1992ki}. In this formalism,
$B_{\phi\rightarrow\pi^0\pi^0\gamma}^\chi=5.1\times 10^{-5}$,
$B_{\phi\rightarrow\pi^0\eta\gamma}^\chi=3.0\times 10^{-5}$,
$B_{\rho\rightarrow\pi^0\pi^0\gamma}^\chi=9.5\times 10^{-6}$,
$B_{\omega\rightarrow\pi^0\pi^0\gamma}^\chi=2.1\times 10^{-7}$,
$B_{\phi\rightarrow K^0\bar K^0\gamma}^\chi=7.6\times 10^{-9}$,
$B_{\rho\rightarrow\pi^0\eta\gamma}^\chi=3.9\times 10^{-11}$ and
$B_{\omega\rightarrow\pi^0\eta\gamma}^\chi=1.5\times 10^{-9}$.
Taking into account these chiral contributions together with the former VMD
contributions, one finally obtains
$B_{\phi\rightarrow\pi^0\pi^0\gamma}^{{\rm VMD}+\chi}=6.1\times 10^{-5}$,
$B_{\phi\rightarrow\pi^0\eta\gamma}^{{\rm VMD}+\chi}=3.6\times 10^{-5}$,
$B_{\rho\rightarrow\pi^0\pi^0\gamma}^{{\rm VMD}+\chi}=2.6\times 10^{-5}$,
$B_{\omega\rightarrow\pi^0\pi^0\gamma}^{{\rm VMD}+\chi}\simeq
 B_{\omega\rightarrow\pi^0\pi^0\gamma}^{\rm VMD}$,
$B_{\phi\rightarrow K^0\bar K^0\gamma}^{{\rm VMD}+\chi}\simeq
 B_{\phi\rightarrow K^0\bar K^0\gamma}^\chi$,
$B_{\rho\rightarrow\pi^0\eta\gamma}^{{\rm VMD}+\chi}\simeq
 B_{\rho\rightarrow\pi^0\eta\gamma}^{\rm VMD}$ and
$B_{\omega\rightarrow\pi^0\eta\gamma}^{{\rm VMD}+\chi}\simeq
 B_{\omega\rightarrow\pi^0\eta\gamma}^{\rm VMD}$,
which, for the first four, are still below the experimental results.
Additional contributions are thus certainly required and the most natural 
candidates are the contributions coming from the exchange of scalar resonances
(as stated in Sect.~\ref{intro}, the contributions from tensor and higher spin
resonances are negligible).
Needless to say that the previous two approaches do not contain the effect
of scalar resonances in an explicit way.
A first model including the scalar resonances explicitly is the
\emph{no structure model}, where the $V\rightarrow P^0P^0\gamma$ decays
proceed through the decay chain $V\rightarrow S\gamma\rightarrow P^0P^0\gamma$,
with $S$ a scalar state, and the coupling $VS\gamma$ is considered as pointlike.
This model seems to be rule out by experimental data on
$\phi\rightarrow\pi^0\pi^0\gamma$ decays \cite{Achasov:2000ym}.
A second model is the well-known \emph{kaon loop model}
(see Ref.~\cite{Achasov:2001cj} and references therein), where the initial vector
decays into a pair of charged pseudoscalar mesons that after the emission of a
photon rescatter into a pair of neutral pseudoscalars through the exchange of
scalar resonances\footnote{It is named the \emph{kaon loop model} because it was
first applied to the $\phi\rightarrow\pi^0\pi^0\gamma$ and
$\phi\rightarrow\pi^0\eta\gamma$ processes \cite{Achasov:1987ts}.}.
In Ref.~\cite{Achasov:1987ts}, it was shown for the first time the convenience of studying the
$\phi\rightarrow P^0P^0\gamma$ processes
in order to investigate the nature of the $f_{0}$ and $a_{0}$ scalar mesons.
In this pioneer work,
$B(\phi\rightarrow f_0\gamma\rightarrow\pi\pi\gamma)=2.5\times 10^{-4}$,
$B(\phi\rightarrow a_0\gamma\rightarrow\pi^0\eta\gamma)=2.0\times 10^{-4}$ and
$B(\phi\rightarrow (f_0+a_0)\gamma\rightarrow K^0\bar K^0\gamma)=1.3\times 10^{-8}$,
for a four-quark structure of the scalar mesons involved, and
$5.4\times 10^{-5}$, $2.4\times 10^{-5}$ and $2.0\times 10^{-9}$, respectively, for a two-quark structure.
In addition, the contribution of the background processes is also calculated,
$B(\phi\rightarrow\rho\pi\rightarrow\pi\pi\gamma)=3.0\times 10^{-5}$ and
$B(\phi\rightarrow\rho^0\pi^0\rightarrow\pi^0\eta\gamma)=0.8\times 10^{-5}$.
Due to its success in predicting the branching ratios as well as the mass spectra,
the \emph{kaon loop model} is used at VEPP-2M and DA$\Phi$NE to extract information on the properties and couplings of the $f_{0}$ and $a_{0}$ from the analysis of experimental data on
$\phi\rightarrow\pi^0\pi^0\gamma$ and $\phi\rightarrow\pi^0\eta\gamma$, respectively
\cite{Achasov:2000ym}--\cite{Aloisio:2002bs}.
A third model that makes use of the characteristics of the two former formalisms is
developed in Refs.~\cite{Gokalp:2003dr}--\cite{Gokalp:2000ir}.
The previous three models include the scalar resonances {\it ad hoc},
{\it i.e.}~the pseudoscalar rescattering amplitudes, where the scalar resonances
are exchanged, are not chiral invariant.
This problem is solved in the next two models which are based not only on the
\emph{kaon loop model} but also on chiral symmetry.
The first one is the Unitarized Chiral Perturbation Theory (UChPT) where the 
scalar resonances do not appear explicitly but are generated dynamically by
unitarizing the one-loop pseudoscalar amplitudes.
In this approach, taking into account only the contribution from unitarized chiral loops,
$B_{\phi\rightarrow\pi^0\pi^0\gamma}^{\rm U\chi PT}=8\times 10^{-5}$,
$B_{\phi\rightarrow\pi^0\eta\gamma}^{\rm U\chi PT}=8.7\times 10^{-5}$,
$B_{\rho\rightarrow\pi^0\pi^0\gamma}^{\rm U\chi PT}=1.5\times 10^{-5}$,
$B_{\omega\rightarrow\pi^0\pi^0\gamma}^{\rm U\chi PT}=4.3\times 10^{-7}$,
$B_{\phi\rightarrow K^0\bar K^0\gamma}^{\rm U\chi PT}=5\times 10^{-8}$,
$B_{\rho\rightarrow\pi^0\eta\gamma}^{\rm U\chi PT}=5.4\times 10^{-11}$ and
$B_{\omega\rightarrow\pi^0\eta\gamma}^{\rm U\chi PT}=2.2\times 10^{-9}$
are obtained \cite{Oller:1998ia,Marco:1999df,Palomar:2001vg}.
A later analysis including more refined production mechanisms gives
$B_{\phi\rightarrow\pi^0\pi^0\gamma}^{\rm U\chi PT}=(1.2\pm 0.3)\times 10^{-4}$ and
$B_{\phi\rightarrow\pi^0\eta\gamma}^{\rm U\chi PT}=(0.6\pm 0.2)\times 10^{-4}$
\cite{Palomar:2003rb}.
The second model is the Linear Sigma Model (L$\sigma$M), a well-defined
$U(3)\times U(3)$ chiral model which incorporates {\it ab initio} both the nonet
of pseudoscalar mesons together with its chiral partner, the scalar mesons nonet.
In this context, the scalar contributions to the $V\rightarrow P^0P^0\gamma$ 
decays are conveniently parametrized in terms of L$\sigma$M amplitudes
compatible with the corresponding ChPT rescattering amplitudes for low dimeson
invariant masses.
The advantage of the L$\sigma$M is to incorporate explicitly the effect of
scalar meson poles while keeping the correct behaviour at low invariant masses
expected from ChPT.

The purpose of the paper is threefold.
Firstly,
to compute the $\phi\to K^0\bar K^0\gamma$ decay
where the scalar effects are known to be dominant.
This process is interesting to study, on one side,
because it allows for a direct measurement of the $K\bar K$ couplings to the $f_0$ and $a_0$  mesons thus avoiding a model dependent extraction
and, on the other side, since it could pose a background problem for testing
CP-violation at DA$\Phi$NE.
The direct measurement of the couplings seems to be feasible in the near future with the
higher luminosity expected at DAFNE-2.
Having 50 fb$^{-1}$, the number of expected $K^0\bar K^0\gamma$ final state is in the
range $2\div 8\times 10^3$ \cite{Ambrosino:2006gk}.
The analysis of CP-violating decays in $\phi\rightarrow K^0\bar K^0$ has been proposed
as a way to measure the ratio $\epsilon^\prime/\epsilon$ \cite{Dunietz:1986jf}, 
but because this means looking for a very small effect,
a $B(\phi\rightarrow K^0\bar K^0\gamma)\gtrsim 10^{-6}$
will limit the precision of such a measurement.
Related to the $\phi\to K^0\bar K^0\gamma$ decay,
there are also the processes $\phi\to (f_0,a_0)\gamma$ which are the main contributions
to the former through the decay chain $\phi\to(f_0+a_0)\gamma\to K^0\bar K^0\gamma$.
An accurate measurement of the production branching ratio and of the mass spectra for
$\phi\to f_0(980)/a_0(980)\gamma$  decays can clarify the controversial nature of these
well established scalar mesons.
Secondly,
to update our previous works on the decays $\phi\to\pi^0\eta\gamma$ \cite{Bramon:2000vu}
and $(\rho, \omega)\to\pi^0\pi^0\gamma$ \cite{Bramon:2001un}
in view of the new experimental data provided by the
KLOE \cite{Aloisio:2002bs}, SND \cite{Achasov:2002jv} and CMD-2 \cite{Akhmetshin:2003rg}
Collaborations.
Concerning $\phi\to\pi^0\eta\gamma$, the update consists in modifying the needed scalar amplitude
to fulfill the chiral constraints at low energy, to incorporate the complete one-loop $a_0$ propagator,
and to show the explicit dependence of that new amplitude on the pseudoscalar mixing angle.
The intermediate vector meson exchange contribution is also included
for comparison with the more precise forthcoming data.
The new analysis of  $(\rho, \omega)\to\pi^0\pi^0\gamma$ includes the most recent values of the
mass and width of the $\sigma$ meson together with a determination of the size of the $f_0$ effects
not considered previously.
Finally,
to calculate for the fist time the scalar contributions to the $\rho\to\pi^0\eta\gamma$ and
$\omega\to\pi^0\eta\gamma$ decays in the framework of the L$\sigma$M where the
scalar meson effects are included in an explicit way.
Our calculation will allow to quantify such contributions with the aim of establishing their relative
impact as compared to the vector meson exchange contributions.
As a matter of completeness, we have also included in this work an updated version of our
former analysis on the $\phi\to\pi^0\pi^0\gamma$ decay \cite{Bramon:2002iw}.

The paper is divided as follows.
In Sect.~\ref{sectChPT}, we discuss and update the contributions from chiral loops
in which our work is inspired.
Sect.~\ref{sectLsM} is devoted to the analysis of the needed scalar amplitudes
in the framework of the L$\sigma$M, showing that these amplitudes can be considered
as improved versions of their chiral counterparts.
The VMD contributions that serve to complete our study are included in
Sect.~\ref{sectVMD}.
The final results where our predictions for the different $V\rightarrow P^0P^0\gamma$
decays are discussed in turn are given in Sect.~\ref{sectResults}.
Finally, in Sect.~\ref{Conclusions} we present our conclusions and comments.
A detailed calculation of the scalar amplitudes as well as
the complete expressions of the one-loop scalar propagators and
the invariant mass distributions are included in Appendix \ref{appendix}.

\section{Chiral-loop predictions}
\label{sectChPT}
The vector meson initiated $V\rightarrow P^0 P^0\gamma$ decays cannot be treated in 
strict Chiral Perturbation Theory (ChPT). 
This theory has to be extended to incorporate on-shell vector meson fields.
At lowest order, this may be easily achieved by means of the ${\cal O}(p^2)$ 
ChPT Lagrangian
\begin{equation}
\label{ChPTLag}
{\cal L}_2=\frac{f^2}{4}\langle D_\mu U^\dagger D^\mu U+M(U+U^\dagger)\rangle\ ,
\end{equation}
where $U=\exp(i\sqrt{2}P/f)$, $P$ is the usual pseudoscalar nonet matrix\footnote{
In order to describe the processes involving the physical $\eta$
in the final state, the singlet term $\eta_0$ has been added to the conventional
octet part.}
and, at this order, $M=\mbox{diag}(m_\pi^2, m_\pi^2, 2m_K^2-m_\pi^2)$ and
$f=f_\pi=92.4$ MeV.
The covariant derivative, now enlarged to include vector mesons, is defined as 
$D_\mu U=\partial_\mu U -i e A_\mu [Q,U] -i g [V_\mu,U]$ with 
$Q=\mbox{diag}(2/3, -1/3, -1/3)$ being the quark charge matrix and $V_\mu$ 
the additional matrix containing the nonet of vector meson fields. 
We follow the conventional normalization for the vector nonet matrix such that the 
diagonal elements are $(\rho^0+\omega_0)/\sqrt{2}, (-\rho^0+\omega_0)/\sqrt{2}$ and
$\phi_0$, where $\omega_0 = (u\bar{u}+d\bar{d})/\sqrt 2$ and $\phi_0 = s\bar{s}$
stand for the ideally mixed states.
The physical $\omega$ and $\phi$ fields are approximately $\omega_0-\epsilon\phi_0$
and $\phi_0+\epsilon\omega_0$ respectively, where
$\epsilon\equiv\varphi_{V}\simeq\sin\varphi_{V}\simeq\tan\varphi_{V}=+0.059\pm 0.004$
accounts for the $\omega$-$\phi$ mixing angle in the flavour basis \cite{Eidelman:2004wy}.

One easily observes that there is no tree-level contribution from the Lagrangian 
(\ref{ChPTLag}) to the $V\rightarrow P^0 P^0\gamma$ amplitudes and that,
at the one-loop level, one needs to compute the set of diagrams shown in 
Ref.~\cite{Bramon:1992ki}.
A straightforward calculation of
$V(q^\ast,\epsilon^\ast)\rightarrow P^0(p)P^0(p^\prime)\gamma(q,\epsilon)$
leads to {\it finite} amplitudes that are conveniently parametrized in the following way:
\begin{eqnarray}
\label{AVpi0pi0ChPT}
{\cal A}_{\rho\rightarrow\pi^0\pi^0\gamma}^{\chi,K} &=&
{\cal A}_{\omega\rightarrow\pi^0\pi^0\gamma}^\chi =
\frac{-1}{\sqrt{2}}{\cal A}_{\phi\rightarrow\pi^0\pi^0\gamma}^\chi=\nonumber\\
&=&\frac{-eg}{2\sqrt{2}\pi^2 m^2_{K^+}}\,\{a\}\,L(m^2_{\pi^0\pi^0})\times
{\cal A}_{K^+ K^-\rightarrow\pi^0\pi^0}^\chi\ ,\\[1ex]
\label{AVpi0etaChPT}
{\cal A}_{\rho\rightarrow\pi^0\eta\gamma}^\chi &=&
{\cal A}_{\omega\rightarrow\pi^0\eta\gamma}^\chi =
\frac{-1}{\sqrt{2}}{\cal A}_{\phi\rightarrow\pi^0\eta\gamma}^\chi=\nonumber\\
&=&\frac{-eg}{2\sqrt{2}\pi^2 m^2_{K^+}}\,\{a\}\,L(m^2_{\pi^0\eta})\times
{\cal A}_{K^+ K^-\rightarrow\pi^0\eta}^\chi\ ,\\[1ex]
\label{Arhopi0pi0ChPTpions}
{\cal A}_{\rho\rightarrow\pi^0\pi^0\gamma}^{\chi,\pi} &=&
\frac{-eg}{\sqrt{2}\pi^2 m^2_{\pi^+}}\,\{a\}\,L(m^2_{\pi^0\pi^0})\times
{\cal A}_{\pi^+\pi^-\rightarrow\pi^0\pi^0}^\chi\ ,\\[1ex]
\label{AphiK0K0barChPT}
{\cal A}_{\phi\rightarrow K^0\bar K^0\gamma}^\chi &=&
\frac{eg}{2\pi^2 m^2_{K^+}}\,\{a\}\,L(m^2_{K^0\bar K^0})\times
{\cal A}_{K^+ K^-\rightarrow K^0\bar K^0}^\chi\ ,
\end{eqnarray} 
where
$\{a\}=(\epsilon^\ast\cdot\epsilon)\,(q^\ast\cdot q)-
       (\epsilon^\ast\cdot q)\,(\epsilon\cdot q^\ast)$
makes the amplitude Lorentz and gauge invariant,
$m^2_{P^0P^0}\equiv s\equiv (p+p^\prime)^2=(q^\ast -q)^2$ is the invariant mass
of the final dimeson system, and $L(m^2_{P^0P^0})$ is the loop integral function
defined as \cite{Bramon:1992ki,Achasov:1987ts,Marco:1999df,Close:ay,LucioMartinez:uw}
\begin{equation}
\label{L}
\begin{array}{rl}
L(m^2_{P^0P^0}) &=\ 
\frac{1}{2(a-b)}-
\frac{2}{(a-b)^2}\left[f\left(\frac{1}{b}\right)-f\left(\frac{1}{a}\right)\right]+\\[2ex]
&+\ \frac{a}{(a-b)^2}\left[g\left(\frac{1}{b}\right)-g\left(\frac{1}{a}\right)\right]\ 
\end{array}
\end{equation}
with 
\begin{equation}
\label{f&g}
\begin{array}{l}
f(z)=\left\{
\begin{array}{ll}
-\left[\arcsin\left(\frac{1}{2\sqrt{z}}\right)\right]^2 & z>\frac{1}{4}\\[1ex]
\frac{1}{4}\left(\log\frac{\eta_+}{\eta_-}-i\pi\right)^2 & z<\frac{1}{4}
\end{array}
\right.\\[5ex]
g(z)=\left\{
\begin{array}{ll}
\sqrt{4z-1}\arcsin\left(\frac{1}{2\sqrt{z}}\right) & z>\frac{1}{4}\\[1ex]
\frac{1}{2}\sqrt{1-4z}\left(\log\frac{\eta_+}{\eta_-}-i\pi\right) & z<\frac{1}{4}
\end{array}
\right.
\end{array}
\end{equation}
and
$\eta_\pm=\frac{1}{2}(1\pm\sqrt{1-4z})$, $a=m^2_V/m^2_{P^+}$ and 
$b=m^2_{P^0P^0}/m^2_{P^+}$.
The coupling constant $g$ comes from the strong amplitude 
${\cal A}(\rho\rightarrow\pi^+\pi^-)=-\sqrt{2}g\,\epsilon^\ast\cdot (p_+-p_-)$  
with $|g|\simeq 4.2$ to agree with 
$\Gamma (\rho\rightarrow\pi^+\pi^-)_{\rm exp}= 150.3$ MeV. 
However, for the $\phi$ decays we replace $g$ by $g_{s}$ where $|g_s|\simeq 4.5$
to agree with $\Gamma (\phi\rightarrow K^+K^-)_{\rm exp}= 2.09$ MeV \cite{Eidelman:2004wy}
---in the good $SU(3)$ limit one should have $|g|=|g_{s}|$.
These couplings are the part beyond standard ChPT which we have fixed
phenomenologically.

As seen from Eqs.~(\ref{AVpi0pi0ChPT}) and (\ref{Arhopi0pi0ChPTpions}),
$\rho\rightarrow\pi^0\pi^0\gamma$ proceeds through a loop of charged kaons or pions,
although the kaon contribution is suppressed by a factor of $10^3$ \cite{Bramon:1992ki}.
$(\omega,\phi)\rightarrow\pi^0\pi^0\gamma$ proceed only by kaon loops since pion loop
contributions are suppressed by $G$-parity for the $\omega$ case and, in addition, by
the Zweig rule for the $\phi$ case
---the same happens for the $\phi\rightarrow K^0\bar K^0\gamma$ case.
$(\rho,\omega,\phi)\rightarrow\pi^0\eta\gamma$ also proceed only by kaon loops because of
isospin conservation.
The four-pseudoscalar amplitudes are found to depend linearly on the variable
$s=m^2_{P^0P^0}$ only\footnote{
The ${\cal A}_{P^+P^-\rightarrow P^0P^0}^\chi$ have not to be understood as
four-pseudoscalar amplitudes in standard ChPT but as terms that factorize from
${\cal A}_{V\rightarrow P^0P^0\gamma}^\chi$ when computed using the chiral-loop
framework of Ref.~\protect\cite{Bramon:1992ki}.}:
\begin{eqnarray}
\label{AKpKmpi0pi0ChPT}
{\cal A}_{K^+K^-\rightarrow\pi^0\pi^0}^\chi &=& \frac{s}{4 f_\pi f_K}\ ,\\[1ex]
\label{AKpKmpi0etaChPT}
{\cal A}_{K^+K^-\rightarrow\pi^0\eta}^\chi &=& \frac{1}{4 f_\pi f_K}
\left[\left(s-\frac{4}{3}m^2_{K}\right)({\rm c}\phi_{P}+\sqrt{2}\,{\rm s}\phi_{P})+
\right.\nonumber\\
&&\left.\qquad\qquad
+\frac{4}{9}(2m^2_{K}+m^2_{\pi})
 \left({\rm c}\phi_{P}-\frac{{\rm s}\phi_{P}}{\sqrt{2}}\right)
\right]\ ,\\[1ex]
\label{Apippimpi0pi0ChPT}
{\cal A}_{\pi^+\pi^-\rightarrow\pi^0\pi^0}^\chi &=& \frac{s-m^2_{\pi}}{f_\pi^2}\ ,\\[1ex]
\label{AKpKmK0K0barChPT}
{\cal A}_{K^+K^-\rightarrow K^0\bar K^0}^\chi &=& \frac{s}{4 f_K^2}\ ,
\end{eqnarray}
where\footnote{Strictly speaking, $f_{K}=f_{\pi}$ at this order in the chiral expansion.}
$f_K= 1.22 f_\pi$, $\phi_P$ is the pseudoscalar mixing angle in the quark-flavour
basis\footnote{
${\cal A}_{K^+K^-\rightarrow\pi^0\eta}^\chi=\frac{1}{\sqrt{6}f_{\pi}^2}
 \left(s-\frac{10}{9}m^2_{K}+\frac{1}{9}m^2_{\pi}\right)$
when the $\eta$-$\eta^\prime$ mixing angle in the octet-singlet basis is fixed to
$\theta_{P}=\arcsin(-1/3)\simeq -19.5^\circ$ as done in Ref.~\protect\cite{Bramon:2000vu}.}
and $({\rm c}\phi_P, {\rm s}\phi_P)\equiv (\cos\phi_P, \sin\phi_P)$.
In this analysis we use $\phi_P=41.8^\circ$ $(\theta_P=-12.9^\circ)$,
a value obtained from the ratio $\phi\to\eta^\prime\gamma/\eta\gamma$
\cite{Aloisio:2002vm} and consistent with a fit to different decay channels \cite{Bramon:1997va}.
It is worth noting that the amplitude in Eq.~(\ref{AKpKmpi0etaChPT}) includes not only
the octet contribution $(\eta_8)$ to the physical $\eta$, as done in Ref.~\cite{Bramon:1992ki},
but also the contribution from the singlet $(\eta_0)$.
In doing so, we have enlarged the initial $SU(3)$-flavour symmetry to $U(3)$ ---nonet symmetry---
in order to obtain the relevant couplings for the singlet,
thus incorporating $\eta$-$\eta^\prime$ mixing effects.
A more general and rigorous extension of ChPT accounting for the effects of the pseudoscalar singlet
in a perturbative way \cite{Kaiser:2000gs}
would coincide at this order with our treatment of the mixing effects.

Integrating the invariant mass distribution for the different $V\rightarrow P^0P^0\gamma$
decays over the whole physical region one obtains the branching ratios shown in
Table \ref{tableChPTvsLsM}.
These results improve the predictions for these processes given in
Ref.~\cite{Bramon:1992ki} where $SU(3)$-breaking effects were ignored
and the pseudoscalar singlet contributions not studied.
\begin{table}
    \centerline{
    \begin{tabular}{ccc}
	\hline\hline\\[-2ex]
	$B(V\rightarrow P^0P^0\gamma)$             & chiral loops               & L$\sigma$M\\[0.5ex]
	\hline\hline\\[-2ex]
	$\phi\rightarrow\pi^0\pi^0\gamma$       & $4.2\times 10^{-5}$   & $1.00\times 10^{-4}$\\[0.5ex]
	$\phi\rightarrow\pi^0\eta\gamma$         & $2.9\times 10^{-5}$   & $7.8\times 10^{-5}$\\[0.5ex]
	$\phi\rightarrow K^0\bar K^0\gamma$  & $4.1\times 10^{-9}$  & $7.5\times 10^{-8}$\\[0.5ex]
	\hline\\[-2ex]
	$\rho\rightarrow\pi^0\pi^0\gamma$       & $1.1\times 10^{-5}$   & $2.2\times 10^{-5}$\\[0.5ex]
	$\rho\rightarrow\pi^0\eta\gamma$         & $6.3\times 10^{-11}$ & $1.3\times 10^{-10}$\\[0.5ex]
	\hline\\[-2ex]
	$\omega\rightarrow\pi^0\pi^0\gamma$ & $1.5\times 10^{-7}$   & $1.7\times 10^{-7}$\\[0.5ex]
	$\omega\rightarrow\pi^0\eta\gamma$   & $1.6\times 10^{-9}$   & $3.4\times 10^{-9}$\\[0.5ex]
	\hline\hline
    \end{tabular}
    }
    \caption{Comparison between chiral-loop and L$\sigma$M predictions of branching ratios
                   for different $V\rightarrow P^0P^0\gamma$ decays.}
    \label{tableChPTvsLsM}
\end{table}

\section{Scalar meson exchange}
\label{sectLsM}
We now turn to the contributions coming from scalar resonance exchange.
From a ChPT perspective their effects are encoded in the low energy constants of the 
higher order pieces of the ChPT Lagrangian.
However, the effects of the scalar states
should manifest in the $V\rightarrow P^0P^0\gamma$ decays not as a constant term
but rather through more complex resonant amplitudes. 
In this section we propose scalar amplitudes which not only obey the ChPT dictates  
in the lowest part of the $P^0P^0$ spectra, as they must,  
but also generate the scalar meson effects for the higher part of the spectra
where the resonant poles should dominate.

The Linear Sigma Model (L$\sigma$M) \cite{Levy:1967,Gasiorowicz:1969kn,Schechter:1971}
will be shown to be particularly appropriate for our purposes. 
In this context, the $V\rightarrow P^0P^0\gamma$ decays proceed through a loop of 
charged pseudoscalar mesons emitted by the initial vector that, after the emission of
a photon (in a gauge-invariant way), can rescatter into $J^{PC}=0^{++}$ pairs of
neutral pseudoscalars. 
The scalar contributions will be conveniently parametrized in terms of L$\sigma$M
amplitudes compatible with ChPT for low dimeson invariant masses.
As an example of our procedure, we discuss in detail the
$K^+ K^-\rightarrow K^0\bar K^0$ rescattering amplitude needed for the 
$\phi\rightarrow K^0\bar K^0\gamma$ process.
The remaining $P^+ P^-\rightarrow P^0 P^0$ rescattering amplitudes are written afterwards.

The $K^+ K^- \rightarrow K^0\bar K^0$ amplitude in the L$\sigma$M turns out to be  
\begin{eqnarray}
\label{AKpKmK0K0barLsM}
{\cal A}_{K^+K^-\rightarrow K^0\bar K^0}^{\mbox{\scriptsize L$\sigma$M}}
&=& g_{K^+ K^- K^0\bar K^0}
    -\frac{g_{\sigma K^+K^-}g_{\sigma K^0\bar K^0}}{s-m^2_\sigma}
    -\frac{g_{f_0 K^+K^-}g_{f_0 K^0\bar K^0}}{s-m^2_{f_0}}\nonumber\\
&&  -\frac{g_{a_0 K^+K^-}g_{a_0 K^0\bar K^0}}{s-m^2_{a_0}}
    -\frac{g^2_{a_{0}^\pm K^\mp K^{0(\bar 0)}}}{t-m^2_{a_{0}}}\ ,
\end{eqnarray}
where the various coupling constants are fixed within the model and can be expressed
in terms of $f_{\pi}$, $f_{K}$,
the masses of the pseudoscalar and scalar mesons involved in the process, and
the scalar meson mixing angle in the flavour basis 
$\phi_S$ \cite{Napsuciale:1998ip,Tornqvist:1999tn}.  
In particular, the $g_{K^+ K^-K^0\bar K^0}$ coupling accounting for the constant
four-pseudoscalar amplitude can be expressed in a more convenient form by imposing that
the ${\cal A}(K^+K^-\rightarrow K^0\bar K^0)_{\mbox{\scriptsize L$\sigma$M}}$
vanishes in the soft-pion(kaon) limit (either $p\rightarrow 0$ or $p'\rightarrow 0$)
---see Appendix \ref{appendix}.  
Then, the amplitude (\ref{AKpKmK0K0barLsM}) can be rewritten as the sum of two terms
each one depending only on $s$ and $t$:
\begin{eqnarray}
\label{AKpKmK0K0barLsMphys}
{\cal A}_{K^+K^-\rightarrow K^0\bar K^0}^{\mbox{\scriptsize L$\sigma$M}}
&\equiv&
{\cal A}_{\mbox{\scriptsize L$\sigma$M}}^s (s)+
{\cal A}_{\mbox{\scriptsize L$\sigma$M}}^t (t)=
\frac{s-m^2_{K}}{4f_K^2}\nonumber\\
&\times &
\left[
\frac{m^2_K-m^2_{\sigma}}{D_{\sigma}(s)}({\rm c}\phi_S -\sqrt{2}{\rm s}\phi_S)^2+
\frac{m^2_K-m^2_{f_0}}{D_{f_0}(s)}({\rm s}\phi_S +\sqrt{2}{\rm c}\phi_S)^2
\right.\nonumber\\
&& \left. -\frac{m^2_K-m^2_{a_0}}{D_{a_0}(s)}\right]
          +\frac{t-m^2_K}{2f_K^2}\frac{m^2_K-m^2_{a_{0}}}{D_{a_{0}}(t)}\ ,
\end{eqnarray}
where $D_{S}(s)$ are the $S=\sigma, f_0, a_0$ propagators   
---similar expressions holds for $D_{a_{0}}(t)$---  
and $({\rm c}\phi_S, {\rm s}\phi_S)\equiv (\cos\phi_S, \sin\phi_S)$.
A Breit-Wigner propagator is used for the $\sigma$, while for the $f_0$ and $a_0$
complete one-loop propagators are preferable (see App.~\ref{appendix})
\cite{Achasov:1987ts,Escribano:2002iv}.

A few remarks on the four-pseudoscalar amplitude in Eq.~(\ref{AKpKmK0K0barLsMphys}) and
its comparison with the chiral-loop amplitude in Eq.~(\ref{AKpKmK0K0barChPT}) are of interest:
\begin{itemize}
\item[\emph{i)}]
for $m_S\rightarrow\infty$ ($S=\sigma, f_0, a_{0}$),
the L$\sigma$M amplitude (\ref{AKpKmK0K0barLsMphys}) reduces to 
\[
\frac{s+t-2m^2_K}{2f_K^2}\ ,
\]
which agrees with the corresponding ChPT amplitude $(2m^2_K-u)/2f_K^2$
if the on-shell condition $s+t+u=4m^2_K$ is invoked. 
This corresponds to the aforementioned complementarity between ChPT and the L$\sigma$M,
thus making the whole analysis quite reliable. 

\item[\emph{ii)}]
the L$\sigma$M amplitude (\ref{AKpKmK0K0barLsMphys}) contains two terms:
${\cal A}^s_{\mbox{\scriptsize L$\sigma$M}}$ and
${\cal A}^t_{\mbox{\scriptsize L$\sigma$M}}$.
The former, being only $s$-dependent, can be directly introduced in
Eq.~(\ref{AphiK0K0barChPT}) instead of ${\cal A}_{K^+ K^-\rightarrow K^0\bar K^0}^\chi$.
Fortunately, this term contains most of the relevant dynamics due to the scalar resonances
---mainly the $f_0$(980) and $a_{0}$(980) poles---
for the process $\phi\rightarrow K^0\bar K^0\gamma$
where the dikaon mass spectrum covers the range $4m_K^2\le m^2_{K\bar K}\le m^2_\phi$.  
The contribution of the remaining term ${\cal A}^t_{\mbox{\scriptsize L$\sigma$M}}$,
being $t$-dependent, would require the calculation of a more involved four-propagator
loop integral.
However, this non-resonant $a_{0}$ contribution can be reasonably estimated by subtracting
from the corresponding chiral-loop amplitude (\ref{AKpKmK0K0barChPT}) the resonant
contributions from the $\sigma$, $f_0$ and $a_{0}$ terms in Eq.~(\ref{AKpKmK0K0barLsMphys}) taking
$m_{\sigma,f_0,a_{0}}\rightarrow\infty$.
The resulting expression corresponds to the desired non-resonant $a_{0}$ contribution
in the $m_{a_{0}}\rightarrow\infty$ limit.
The difference between this $a_{0}$ contribution and the same contribution for an $a_{0}$
of finite mass can be considered as negligible due to the high mass of the $a_{0}$ and the
lack of poles in the $t$-channel. 
Therefore, an improved expression for the chiral-loop amplitude in
Eq.~(\ref{AKpKmK0K0barChPT}) taking into account the (pole dominated) $s$-channel scalar 
dynamics is the following:  
\begin{eqnarray}
\label{AKpKmK0K0barChPTLsM}
{\cal A}_{K^+K^-\rightarrow K^0\bar K^0}^{\mbox{\scriptsize L$\sigma$M}}
&=&
\frac{s-m^2_{K}}{4f_K^2}\left[
\frac{m^2_K-m^2_{\sigma}}{D_{\sigma}(s)}({\rm c}\phi_S -\sqrt{2}{\rm s}\phi_S)^2
\right.\nonumber\\
&& \left.
   +\frac{m^2_K-m^2_{f_0}}{D_{f_0}(s)}({\rm s}\phi_S +\sqrt{2}{\rm c}\phi_S)^2
   -\frac{m^2_K-m^2_{a_0}}{D_{a_0}(s)}\right]\nonumber\\
&& +\frac{m^2_{K}-s/2}{2f_K^2}\ .
\end{eqnarray}
As desired, it has no $t$ dependence and can thus be plugged in Eq.~(\ref{AphiK0K0barChPT})
instead of ${\cal A}_{K^+K^-\rightarrow K^0\bar K^0}^\chi$.
One then obtains the $s$-dependent amplitude  
\begin{equation}
\label{AphiK0K0barLsM}
{\cal A}_{\phi\rightarrow K^0\bar K^0\gamma}^{\mbox{\scriptsize L$\sigma$M}}= 
\frac{eg_{s}}{2\pi^2 m^2_{K^+}}\,\{a\}\,L(s)\times
{\cal A}_{K^+ K^-\rightarrow K^0\bar K^0}^{\mbox{\scriptsize L$\sigma$M}} \ ,
\end{equation}
which will be used from now on.  
Notice that for large scalar masses one recovers Eq.~(\ref{AphiK0K0barChPT}), 
that the $f_0(980)$, $a_0(980)$ (and, eventually, $\sigma(600)$) $s$-channel poles
do now appear, and that the remaining term, $(m^2_{K}-s/2)/2f_K^2$, accounts for $a_{0}$
exchange effects in the $t$-channel.

\item[\emph{iii)}]
the large widths of the scalar resonances break chiral symmetry if they are naively
introduced in amplitudes, an effect already noticed in Ref.~\cite{Achasov:1994iu}.
Accordingly, we introduce the $f_0(980)$ and $a_0(980)$ widths in the propagators
only {\it after} chiral cancellation of constant terms in the amplitude (\ref{AKpKmK0K0barLsMphys}).
In this way the pseudo-Goldstone nature of kaons is preserved.

\item[\emph{iv)}]
the $K^0\bar K^0$ invariant mass spectrum for the
$\phi\rightarrow K^0\bar K^0\gamma$ decay covers the region where the presence of the
$f_0(980)$ and the $a_0(980)$ mesons should manifest.
Because of the corresponding propagators in Eq.~(\ref{AKpKmK0K0barChPTLsM}),
one should also be able to reproduce the effects of the $f_0$ and the $a_0$ poles
at all $K^0\bar K^0$ invariant mass values.
\end{itemize}

Following the same procedure for the rest of relevant four-pseudoscalar amplitudes,
one gets
\begin{eqnarray}
\label{AKpKmpi0pi0ChPTLsM}
{\cal A}_{K^+K^-\rightarrow\pi^0\pi^0}^{\mbox{\scriptsize L$\sigma$M}} &=&
\frac{s-m^2_{\pi}}{2f_\pi f_K}
\left[\frac{m^2_K-m^2_{\sigma}}{D_{\sigma}(s)}\,
{\rm c}\phi_S({\rm c}\phi_S-\sqrt{2}\,{\rm s}\phi_S)+\right.\nonumber\\
&&\left.+
\frac{m^2_K-m^2_{f_0}}{D_{f_0}(s)}\,
{\rm s}\phi_S({\rm s}\phi_S+\sqrt{2}\,{\rm c}\phi_S)
\right]
+\frac{m^2_{\pi}-s/2}{2f_\pi f_K}\ ,\\[1ex]
\label{AKpKmpi0etaChPTLsM}
{\cal A}_{K^+K^-\rightarrow\pi^0\eta}^{\mbox{\scriptsize L$\sigma$M}} &=&
\frac{s-m^2_{\eta}}{2f_\pi f_K}
\frac{m^2_K-m^2_{a_0}}{D_{a_0}(s)}\,{\rm c}\phi_{P}+\nonumber\\
&&+\frac{m^2_{\eta}+m^2_\pi-s}{4f_\pi f_K}({\rm c}\phi_{P}-\sqrt{2}\,{\rm s}\phi_{P})\ ,\\[1ex]
\label{Apippimpi0pi0ChPTLsM}
{\cal A}_{\pi^+\pi^-\rightarrow\pi^0\pi^0}^{\mbox{\scriptsize L$\sigma$M}} &=&
\frac{s-m^2_\pi}{f_\pi^2}
\left(\frac{m^2_\pi-m^2_\sigma}{D_\sigma(s)}\,{\rm c}^2\phi_S+ 
      \frac{m^2_\pi-m^2_{f_0}}{D_{f_0}(s)}\,{\rm s}^2\phi_S\right)\ .
\end{eqnarray}
As stated before, these amplitudes not only are compatible with their chiral-loop counterparts
for low dimeson invariant masses but also generate the scalar poles that are dominant for higher
invariant masses.
The amplitudes (\ref{AKpKmK0K0barChPTLsM}) and
(\ref{AKpKmpi0pi0ChPTLsM})--(\ref{Apippimpi0pi0ChPTLsM}) are a clear manifestation of the
pursued complementarity between ChPT and the L$\sigma$M.
However, Eq.~(\ref{AKpKmpi0etaChPTLsM}) is an exception.
When all the scalar masses are sent to infinity the resulting amplitude does not match
the chiral-loop one.
This is because of the different treatment of $SU(3)$-breaking corrections in ChPT and the L$\sigma$M.
In Eq.(\ref{AKpKmpi0etaChPT}), the Gell-Mann--Okubo mass relation
$3m^2_{\eta_8}-4m^2_K+m^2_\pi=0$
has been used to express the amplitude in terms of $m^2_K$ and $m^2_\pi$,
while in the L$\sigma$M this relation is violated by second order $SU(3)$-breaking corrections
and therefore only satisfied in the exact $SU(3)$ limit.
Consequently, the amplitudes (\ref{AKpKmpi0etaChPT}) and (\ref{AKpKmpi0etaChPTLsM})
can only be compared in the infinite scalar mass limit once the $SU(3)$ limit has been taken previously.
In that case, the pseudoscalar singlet is decoupled from the octet since $\theta_P=0$
(which implies $\phi_P=\arctan\sqrt{2}$)
and the whole pseudoscalar octet have a common mass,
thus making both amplitudes to coincide in the $m_S\to\infty$ limit.

The final amplitudes
${\cal A}_{V\rightarrow P^0P^0\gamma}^{\mbox{\scriptsize L$\sigma$M}}$
are obtained replacing 
${\cal A}_{P^+P^-\rightarrow P^0P^0}^{\chi}$ from the chiral-loop expressions in
Eqs.~(\ref{AVpi0pi0ChPT}--\ref{AphiK0K0barChPT})
by the corresponding L$\sigma$M counterparts
${\cal A}_{P^+P^-\rightarrow P^0P^0}^{\mbox{\scriptsize L$\sigma$M}}$ shown above.
Integrating the invariant mass spectrum for the different $V\rightarrow P^0P^0\gamma$
processes one finally obtains the branching ratios shown in Table \ref{tableChPTvsLsM}.
As seen, taking into account the effect of scalar poles makes the L$\sigma$M predictions
bigger than the chiral-loop predictions.

\section{Vector exchange}
\label{sectVMD}
In addition to the L$\sigma$M contributions, which can be viewed as an
improved version of the chiral-loop predictions, the analysis should be
extended to include vector meson exchange in the $t$- and $u$-channel. 
These VMD contributions were already considered in Ref.~\cite{Bramon:1992kr}. 
In this framework, $V^0\rightarrow P^0P^0\gamma$ proceed through the exchange of
intermediate (virtual) vector mesons mesons ($V$ and $V^\prime$) in the direct
and crossed channels of the  
$V^0\rightarrow (V, V^\prime) P^0\rightarrow\ P^0P^0\gamma$ decay chain.

In order to describe these vector meson contributions we use the $SU(3)$-symmetric
Lagrangians
\begin{equation}
\label{VMDLag}
\begin{array}{rcl}
{\cal L}_{\rm VVP}&=&\frac{G}{\sqrt{2}}\epsilon^{\mu\nu\alpha\beta}
                     \langle\partial_\mu V_\nu\partial_\alpha V_\beta P\rangle\ ,\\[2ex]
{\cal L}_{{\rm V}\gamma}&=& -4f^2 e g A_\mu\langle Q V^\mu\rangle\ ,
\end{array}
\end{equation}
where $G=\frac{3g^2}{4\pi^2 f}$ is the $\omega\rho\pi$ coupling constant, $f=f_{\pi}$,
and $|g|\simeq 4.0$ as follows from various $\rho$ and $\omega$ decay data
\cite{Eidelman:2004wy,daphne95:bramon}.
The VMD amplitude for
$V(q^\ast,\epsilon^\ast)\rightarrow P^0(p)P^0(p^\prime)\gamma(q,\epsilon)$
is then found to be  
\begin{equation}
\label{AVVMD}
\textstyle
{\cal A}_{V\rightarrow P^0P^0\gamma}^{\rm VMD}=
C(V^0P^0P^0\gamma)\frac{G^2 e}{\sqrt{2}g}
\left(\frac{P^2\{a\}+\{b(P)\}}{M^2_{V}-P^2-i M_{V}\Gamma_{V}}+
      \frac{{P^\prime}^2\{a\}+\{b({P^\prime})\}}
           {M^2_{V^\prime}-P^2-i M_{V\prime}\Gamma_{V\prime}}\right)\ ,
\end{equation}
with $\{a\}$ as in Eq.~(\ref{AVpi0pi0ChPT}) and a new amplitude 
\begin{equation}
\label{b}
\begin{array}{rcl}
\{b(P)\}&=& -(\epsilon^\ast\cdot\epsilon)\,(q^\ast\cdot P)\,(q\cdot P)
            -(\epsilon^\ast\cdot P)\,(\epsilon\cdot P)\,(q^\ast\cdot q)\\[1ex]
        & & +(\epsilon^\ast\cdot q)\,(\epsilon\cdot P)\,(q^\ast\cdot P)
            +(\epsilon\cdot q^\ast)\,(\epsilon^\ast\cdot P)\,(q\cdot P)\ ,
\end{array}
\end{equation}
where $P=p+q$ and $P^\prime=p^\prime+q$ are the momenta of the intermediate
$V$ and $V^\prime$ mesons in the $t$- and $u$-channel, respectively.
The intermediate vector mesons, $V$ and $V^\prime$,
can be either the $\omega$ or the $\rho$ mesons, with $V=V^\prime$ in $\pi^0\pi^0\gamma$
and  $V\neq V^\prime$ in $\pi^0\eta\gamma$;
for $\phi\rightarrow K^0\bar K^0\gamma$ one obviously has $V=K^{\ast 0}$ and
$V^\prime=\bar K^{\ast 0}$.
The coefficient $C$ is the same for both terms (using $SU(3)$-symmetric couplings)
and changes from process to process according to well-known quark-model or
nonet-symmetry rules:
\begin{eqnarray}
\label{C}
1=C(\rho\rightarrow\pi^0\pi^0\gamma)=3C(\omega\rightarrow\pi^0\pi^0\gamma)=
            -\frac{3}{\sqrt{2}}C(\phi\rightarrow K^0\bar K^0\gamma)\ ,\\
\cos\phi_P =3C(\rho\rightarrow\pi^0\eta\gamma)=C(\omega\rightarrow\pi^0\eta\gamma)\ ,
\label{pi0etagammacoupling}
\end{eqnarray}
and
\begin{equation}
\label{Ceps}
\epsilon=3C(\phi\rightarrow\pi^0\pi^0\gamma)=
                 \sec\phi_P\, C(\phi\rightarrow\pi^0\eta\gamma)\ ,
\end{equation}
for the $\phi$ decays where the Zweig rule is operative.
Notice the explicit dependence on the mixing angle for the processes involving the
$\eta$ meson in the final state.
Further details on these contributions are given in the following section. 

\section{Results}
\label{sectResults}
The final amplitudes ${\cal A}(V\rightarrow P^0P^0\gamma)$  
are thus the sum of the VMD contribution in Eq.~(\ref{AVVMD}) plus the L$\sigma$M  
contribution containing the scalar resonance effects in Eq.~(\ref{AKpKmK0K0barChPTLsM}) and 
Eqs.~(\ref{AKpKmpi0pi0ChPTLsM}--\ref{Apippimpi0pi0ChPTLsM}).
The $P^0P^0$ invariant mass distribution for these processes is given in Appendix \ref{invmassdis}.
The results of our analysis are now discuss in turn for each process.

\subsection{$\phi\rightarrow K^0\bar K^0\gamma$}
As stated in the Introduction,
this process is the only radiative decay where the $f_0$ and $a_0$ scalar mesons
contribute simultaneously.
This will allow, once this decay is measured presumably at DAFNE-2,
to extract relevant information on the nature and couplings of both mesons,
and to compare it with the one obtained from the already experimentally measured
$\phi\rightarrow\pi^0\pi^0\gamma$ and $\phi\rightarrow\pi^0\eta\gamma$ decays.
Therefore, a theoretical prediction for the different contributions to the
branching ratio and mass spectrum of this process is welcome and useful.
In addition, our analysis will serve to certify that  the decay $\phi\rightarrow K^0\bar K^0\gamma$
cannot pose a background problem for testing CP-violation at DA$\Phi$NE
since the calculated branching ratio is well below $10^{-6}$ (see below).

\begin{figure}[t]
\centerline{\includegraphics[width=0.85\textwidth]{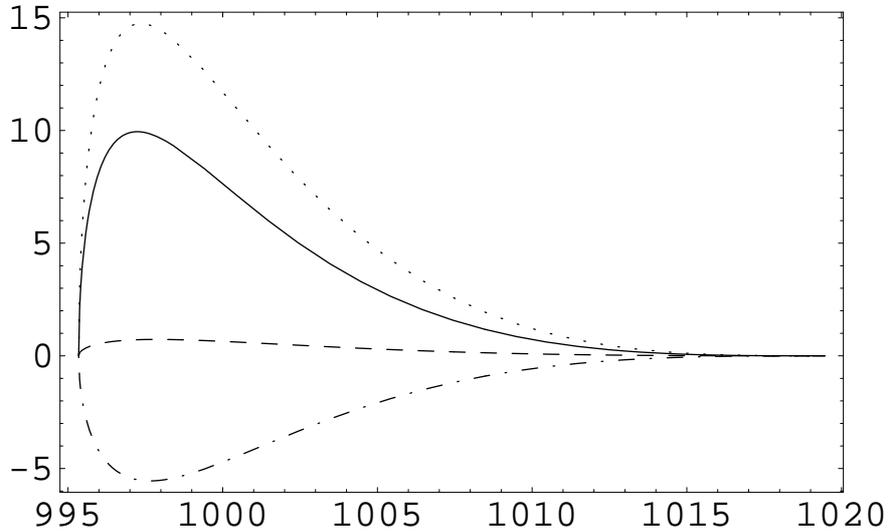}}  
\caption{\small
$dB(\phi\rightarrow K^0\bar K^0\gamma)/dm_{K^0\bar K^0}\times 10^9$ (in units of MeV$^{-1}$)   
as a function of the dikaon invariant mass $m_{K^0\bar K^0}$ (in MeV).
The dotted, dashed and dot-dashed lines correspond to the separate contributions 
from $f_0$, $a_0$ and their interference, respectively. 
The solid line is the total result.}
\label{dBdmKKgamma}
\end{figure}
The separate contributions from $f_0$, $a_0$ and their interference,  
as well as the total result are shown in Fig.~\ref{dBdmKKgamma}.
These mass spectra are computed assuming masses for the scalar resonances of
$m_{f_0}=985$ MeV and $m_{a_0}=984.7$ MeV,
and a pseudoscalar (scalar) mixing angle of $\phi_P=41.8^\circ$ ($\phi_S=-8^\circ$).
The values of the $f_0$ mass and the scalar mixing angle are obtained from the
$\phi\rightarrow\pi^0\pi^0\gamma$ analysis (see the dedicated subsection)
whereas the $a_0$ mass is taken from Ref.~\cite{Eidelman:2004wy} and
$\phi_P$ from the ratio $\phi\to\eta^\prime\gamma/\eta\gamma$ \cite{Aloisio:2002vm}.
As seen, the $f_0$ contributes more strongly than the $a_0$
due to a smaller imaginary part of the propagator and a larger coupling to kaons.
The interference is negative since isospin invariance implies
$g_{f_0K^+K^-}=g_{f_0K^0\bar K^0}$ and $g_{a_0K^+K^-}=-g_{a_0K^0\bar K^0}$.
Integrating the $K^0\bar K^0$ invariant mass spectrum one obtains for the scalar contribution
$B(\phi\to K^0\bar K^0\gamma)_{\mbox{\scriptsize L$\sigma$M}}=7.5\times 10^{-8}$.
It is worth mentioning that the value obtained is very sensitive to the scalar mixing angle,
a change of $1^\circ$ modifies the branching ratio around 20\%.
The whole scalar contribution includes not only the $f_0$ and $a_0$ effects but also the $\sigma$ ones.
However, the latter are negligible due to the suppression of the $\sigma K\bar K$ coupling if\footnote{
A value of $m_{\sigma}=478$ MeV is taken from the analysis of $D^+\to\pi^-\pi^+\pi^+$ performed
by the E791 Coll.~\cite{Aitala:2000xu} at Fermilab.}
$m_\sigma\simeq m_K$ and for kinematical reasons.
Numerically, they amount to less than 1\% of the total result.
Finally, the exchange of intermediate $K^{\ast 0}$ and $\bar K^{\ast 0}$
vector mesons is calculated to be
$B(\phi\to K^0\bar K^0\gamma)_{\rm VMD}=2.0\times 10^{-12}$ and therefore negligible.
In summary, our predicted branching ratio including the scalar resonances explicitly is
$B(\phi\to K^0\bar K^0\gamma)=7.5\times 10^{-8}$.
This value is in agreement with previous phenomenological
estimates \cite{Oller:1998ia,Close:ay,Achasov:2001rn}
(see also Ref.~\cite{Close:ay} for a review of earlier predictions).
Notice that the branching ratio obtained here is one order of magnitude larger than the
chiral-loop prediction $B(\phi\to K^0\bar K^0\gamma)=4.1\times 10^{-9}$.
However, it is still one order of magnitude smaller than the limit, ${\cal O}(10^{-6})$,
in order to pose a background problem for testing CP-violating decays at DA$\Phi$NE.

\subsection{$\phi\rightarrow f_{0}\gamma/a_{0}\gamma$}
\label{ratio}
In the kaon loop model, these two processes are driven by the decay 
chain
$\phi\rightarrow K^+K^-(\gamma)\rightarrow f_{0}\gamma$ and
$a_{0}\gamma$.
The amplitudes are given by
\begin{equation}
\label{Aratio}
{\cal A}=\frac{eg_{s}}{2\pi^2 m^2_{K^+}}\,\{a\}\,L(m^2_{f_{0}(a_{0})})
\times g_{f_{0}(a_{0})K^+K^-}\, ,
\end{equation} 
where the scalar coupling constants are fixed within the L$\sigma$M to
\begin{equation}
\label{couplingsratio}
g_{f_{0}K^+K^-}=\frac{m^2_{K}-m^2_{f_{0}}}{2f_{K}}
({\rm s}\phi_S +\sqrt{2}{\rm c}\phi_S)\, ,\quad
g_{a_{0}K^+K^-}=\frac{m^2_{K}-m^2_{a_{0}}}{2f_{K}}\, .
\end{equation}
The ratio of the two branching ratios is thus
\begin{equation}
\label{R}
R_{\phi\rightarrow f_{0}\gamma/a_{0}\gamma}^{\mbox{\scriptsize L$\sigma$M}}=
\frac{|L(m^2_{f_{0}})|^2}{|L(m^2_{a_{0}})|^2}
\frac{\left(1-m^2_{f_{0}}/m^2_{\phi}\right)^3}
       {\left(1-m^2_{a_{0}}/m^2_{\phi}\right)^3}
       \times\frac{g^2_{f_{0}K^+K^-}}{g^2_{a_{0}K^+K^-}}
\simeq ({\rm s}\phi_S +\sqrt{2}{\rm c}\phi_S)^2\, ,
\end{equation}
where the approximation is valid for $m_{f_{0}}\simeq m_{a_{0}}$.
For $\phi_{S}=-8^\circ$, one gets
$R_{\phi\rightarrow f_{0}\gamma/a_{0}\gamma}^{\mbox{\scriptsize L$\sigma$M}} \simeq 1.6$
which should be compared with the experimental value
$R_{\phi\rightarrow f_{0}\gamma/a_{0}\gamma}^{\mbox{\scriptsize KLOE}}
 =6.1\pm 0.6$ \cite{Aloisio:2002bs}.
However, this value is obtained from a large destructive interference 
between the $f_{0}\gamma$ and $\sigma\gamma$ contributions to
$\phi\rightarrow\pi^0\pi^0\gamma$, in disagreement with other 
experiments \cite{Achasov:2000ym}.
In addition, the approximate expression for the ratio (\ref{R}) is only valid when the $f_0$ mass
is below the charged kaon threshold ($2m_{K^+}\simeq 987$ MeV).
If not, the steep behaviour of the loop function after threshold makes the approximation meaningless
and the exact expression has to be taken.
In this case, $R_{\phi\rightarrow f_{0}\gamma/a_{0}\gamma}^{\mbox{\scriptsize L$\sigma$M}}$
can be much smaller than the given prediction (independently of the mixing angle value).
Conversely, if the $a_0$ mass, which we have kept fixed to $m_{a_0}=984.7$ MeV,
is moved to a value above threshold, then the ratio can be even larger than the experimental result.
Therefore, a precise fit of the $f_0$ and $a_0$ masses using the 
$\phi\to\pi^0\pi^0\gamma$ and $\pi^0\eta\gamma$ decays
is mandatory before drawing definite conclusions on $R_{\phi\rightarrow f_{0}\gamma/a_{0}\gamma}$.
Once these masses are fixed and assuming
the validity of the kaon loop model as the production mechanism and 
the L$\sigma$M as the model for fixing the $SPP$ couplings,
the measurement of this ratio could be used to get some insight into the value of the scalar mixing angle.

\subsection{$\phi\rightarrow\pi^0\pi^0\gamma$}
\label{phipi0pi0gamma}
This process is considered the first of the so-called \emph{golden processes}
since the $f_0$ contribution is seen to be dominant.
Hence, the precise measurements of the total branching ratio and mass spectrum
performed at SND \cite{Achasov:2000ym}, CMD-2 \cite{Akhmetshin:1999di} and
KLOE \cite{Aloisio:2002bt}
allow to extract valuable information on the properties and couplings of this resonance.
In this analysis, we predict the scalar effects on this interesting and important process
using the L$\sigma$M as the framework for calculating the $f_0$ and $\sigma$ resonant contributions.
In addition, we also predict the vector meson exchange effects.

\begin{figure}[t]
\centerline{\includegraphics[width=0.85\textwidth]{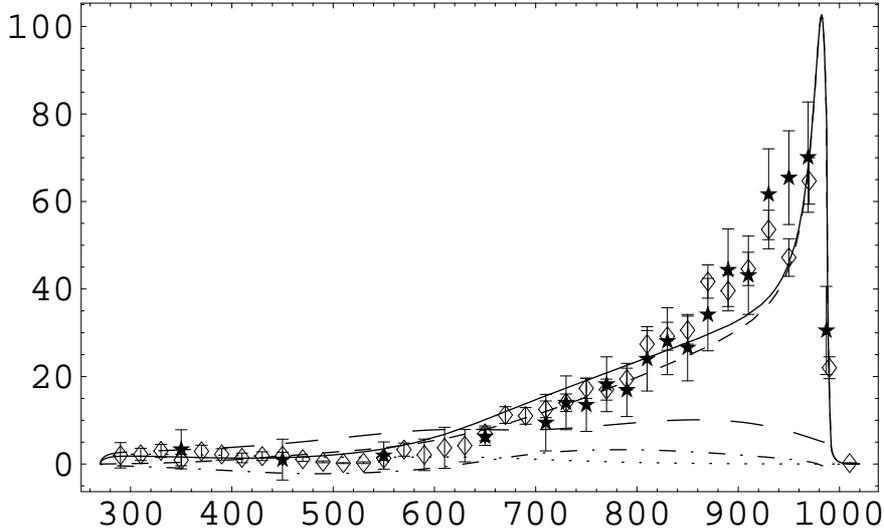}}  
\caption{\small
$dB(\phi\rightarrow\pi^0\pi^0\gamma)/dm_{\pi^0\pi^0}\times 10^8$ (in units of MeV$^{-1}$)   
as a function of the dipion invariant mass $m_{\pi^0\pi^0}$ (in MeV).
The dotted, dashed and dot-dashed lines correspond to the separate contributions 
from VMD, L$\sigma$M and their interference, respectively. 
The solid line is the total result. 
The long-dashed line is the chiral-loop prediction.
Experimental data are taken from Ref.~\cite{Achasov:2000ym} (solid star) and
Ref.~\cite{Aloisio:2002bt} (open diamond).}
\label{dBdmphipi0pi0gamma}
\end{figure}
The $\pi^0\pi^0$ invariant mass distribution, with the separate contributions
from the L$\sigma$M, VMD and their interference, as well as the total result, are shown in
Fig.~\ref{dBdmphipi0pi0gamma}.
We use $m_\sigma=478$ MeV \cite{Aitala:2000xu}, $\Gamma_\sigma=258$ MeV,
as required by the L$\sigma$M, $m_{f_0}=985$ MeV and $\phi_S=-8^\circ$.
The latter two values are obtained from the best fit to experimental data in
Refs.~\cite{Achasov:2000ym,Aloisio:2002bt}
and are the only two free parameters of our prediction.
The global agreement with the data is rather acceptable.
As expected, $f_0(980)$ scalar meson exchange contributes decisively to achieve this agreement.
Indeed, for the integrated branching ratio one obtains
$B(\phi\to\pi^0\pi^0\gamma)_{\mbox{\scriptsize L$\sigma$M+VMD}}=1.11\times 10^{-4}$,
quite in line with the experimental results quoted in Refs.~\cite{Achasov:2000ym,Aloisio:2002bt}
and previous analyses \cite{Achasov:2001cj,Gokalp:2001bk,Palomar:2003rb}.
Without the VMD contribution, our result decreases only by some 10\%,
$B(\phi\to\pi^0\pi^0\gamma)_{\mbox{\scriptsize L$\sigma$M}}=1.00\times 10^{-4}$,
but it still remains well above the chiral-loop value,
$B(\phi\to\pi^0\pi^0\gamma)_{\chi}=4.2\times 10^{-5}$, 
which did not contain the crucial scalar contributions.

The reason for not achieving a better accord between our prediction for the $\pi^0\pi^0$ mass spectrum and the KLOE experimental data is the simplicity of our signal amplitude.
It is derived from a tree level L$\sigma$M where the scalar contributions to the $s$-channel,
the dominant production mechanism, have been kept explicitly while other contributions, scalar or not, have been integrated out in a manner that preserves the chiral-loop result
(see App.~\ref{scalaramp} for details).
The advantage of using such a simple amplitude is its predictivity, \textit{i.e.}~with only two parameters (the couplings of $\sigma$ and $f_0$ to pions and kaons are fixed within the model and the $\sigma$ mass is fixed from experiment)
one is able to get a reasonable description of the mass spectrum and a better value for the integrated branching ratio.
Besides, our background amplitude is deduced from a VMD model where the required $\phi\rho\pi^0$ and $\rho\pi^0\gamma$ couplings are written in terms of a single parameter $g$ which is also fixed from experiment.
A way of improving our amplitude could be on the side of the background to include an additional phase to take into account $\rho\pi$ scattering effects and fix $\phi\rho\pi^0$ and $\rho\pi^0\gamma$ independently from experiment.
On the side of the signal, it would be welcome to incorporate next-to-leading effects in the mixing matrix of  $f_0$ and $\sigma$, beyond the tree level description in terms of a mixing angle, and include more decay channels in the propagators.
In this direction, an interesting analysis is that of Ref.~\cite{Achasov:2005hm} where a more involved amplitude also inspired by the L$\sigma$M is used for describing simultaneously $\phi\to\pi^0\pi^0\gamma$ and $\pi\pi$ scattering.
Their amplitudes accommodate all the above improvements as well as a phase for the elastic background of the $\pi\pi$ and $K\bar K$ scattering.
As a result, the agreement between their predictions and the KLOE experimental data is significantly better than ours.
Nevertheless, the values for the scalar masses and the $\sigma$ width they obtained from their fits are close to ours (see Ref.~\cite{Achasov:2005hm}) and the hierarchy of the scalar couplings follows the pattern dictated by the L$\sigma$M, thus giving more consistency to our simpler approach.

\begin{figure}[t]
\centerline{\includegraphics[width=1.05\textwidth]{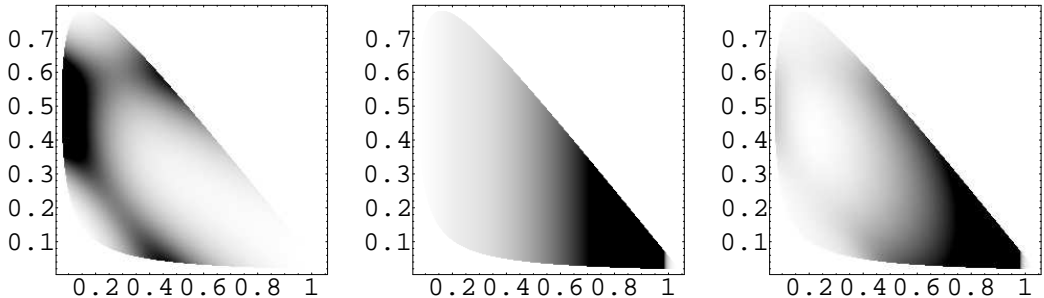}}  
\caption{\small
$dB(\phi\rightarrow\pi^0\pi^0\gamma)/dm^2_{\pi^0\pi^0}dm^2_{\pi^0\gamma}$ (in units of GeV$^{-4}$)   
as a function of $m^2_{\pi^0\pi^0}$ (in GeV$^2$) ---\emph{horizontal axis}---
and $m^2_{\pi^0\gamma}$ (in GeV$^2$) ---\emph{vertical axis}---.
The left, centered and right Dalitz plots correspond to the VMD, L$\sigma$M and total contributions,
respectively.}
\label{DPphipi0pi0gamma}
\end{figure}

It is worth noting that our prediction is very sensitive to the $f_0$ mass and scalar mixing angle.
As an example, 
$B(\phi\to\pi^0\pi^0\gamma)=(1.20,0.98)\times 10^{-4}$ for $m_{f_0}=(980,990)$ MeV and
$B(\phi\to\pi^0\pi^0\gamma)=(1.16,1.06)\times 10^{-4}$ for $\phi_S=(-9^\circ,-7^\circ)$.
The dependence on both parameters is because of the $f_0\pi\pi$ coupling
which in the L$\sigma$M is proportional to $m^2_{f_0}$ and $\sin\phi_S$.
This coupling enters into the calculation not only in the scalar part of the amplitude but also
on the imaginary part of the propagator
(remember that we use a complete one-loop propagator for this resonance).
In addition, if the $f_0$ mass is above the two charged kaon threshold
the kaon-loop function drops drastically thus decreasing the $f_0$ contribution.
Consequently, the $\phi\to\pi^0\pi^0\gamma$ decay could be used to obtain relevant information
on these two parameters.
Concerning the $\sigma$ contribution,
it is suppressed with respect to the $f_0$,
since $g_{\sigma K\bar K}\propto (m^2_\sigma-m^2_K)\simeq 0$ for $m_\sigma\simeq m_K$,
but it still amounts to 10\% of the total result.
By contrast, the chiral-loop prediction shows no suppression in the region $m_{\pi\pi}\simeq 500$ MeV,
see Fig.~\ref{dBdmphipi0pi0gamma}.
The effects of the $\sigma$ are enhanced
if instead of using $m_\sigma=478$ MeV,
one takes $m_\sigma=513$ MeV from the Dalitz analysis of $D^0\to K_S^0\pi^+\pi^-$ by the
CLEO Coll.~\cite{Muramatsu:2002jp}.
In this case, however, the fit to the $\pi^0\pi^0$ mass spectrum is worse and
the branching ratio incompatible with the KLOE result.
On the contrary, our prediction is hardly sensitive to the $\sigma$ decay width.
Changing from the L$\sigma$M value $\Gamma_\sigma=258$ MeV for $m_\sigma=478$ MeV
to $\Gamma_\sigma=324$ MeV \cite{Aitala:2000xu}
or $\Gamma_\sigma=335$ MeV \cite{Muramatsu:2002jp} makes no substantial difference
(less than 1\% in both cases).
Therefore, the dependence on the $\sigma$ parameters, particularly on $m_\sigma$,
could be used in the near future to fit its value once more precise experimental data on
$\phi\to\pi^0\pi^0\gamma$ is available.
Finally, the VMD contribution is calculated to be
$B(\phi\to\pi^0\pi^0\gamma)_{\rm VMD}=8.3\times 10^{-6}$
that although small should not be considered as negligible since as stated before it is responsible
together with its interference with the scalar part for a 10\% of the final result.
For the sake of comparison with experiment, we also show in Fig.~\ref{DPphipi0pi0gamma}
the predicted Dalitz plots for the scalar and vector meson exchange contributions,
as well as the total result.
As seen, the VMD plot shows the exchange of intermediate $\rho$ mesons at
$m^2_{\pi^0\gamma}\simeq m^2_\rho$ (horizontal band) and
$m^2_{\pi^0\gamma}+m^2_{\pi^0\pi^0}\simeq 2m^2_{\pi^0}+m^2_\phi-m^2_\rho$ (inclined band),
while the L$\sigma$M plot shows the exchange of an $f_0(980)$ at 
$m^2_{\pi^0\pi^0}\simeq m^2_{f_0}$ (vertical band).

One way to further test the goodness of our approach applied to this decay, in particular for the amplitude used, would be to perform a simultaneous analysis of
$\phi\to\pi^0\pi^0\gamma$ and $\pi\pi$ scattering along the lines of Ref.~\cite{Achasov:1997ih}.
By unitarity, the phase associated to the $I=J=0$ partial wave of the $K^+K^-\to\pi^0\pi^0$ amplitude must be equal below the kaon threshold to the phase $\delta_0^0(\pi\pi\to\pi\pi)$.
This kind of analysis, which is beyond the scope of the present one and left for future work, has been already carried out for the SND data in Ref.~\cite{Achasov:2001cj} and for the more precise KLOE data in Ref.~\cite{Achasov:2005hm},
showing in both cases that with an amplitude more elaborate than ours (see previous discussion) it is possible to get simultaneously a remarkable fit to the $\pi^0\pi^0$ mass spectrum and the $\pi\pi$ phaseshift.

\subsection{$\phi\rightarrow\pi^0\eta\gamma$}
\label{phipi0etagamma}
The second of the \emph{golden processes}, $\phi\rightarrow\pi^0\eta\gamma$,
is interesting to analyze since it provides relevant information on the properties and couplings of the
$a_0$ resonance, whose effects on the process are also seen to be experimentally dominant.
Here, we present a parameter-free prediction for the different contributions, scalar and vector,
to this decay using the L$\sigma$M for calculating the $K\bar K\to\pi\eta$ invariant amplitude.
Our calculation of the scalar contribution not only satisfies the chiral constraints at low energy
but also incorporates the pole effects of the $a_0$ in an explicit way.
In this manner, we are able to nicely reproduce the peak and the lower tail of the mass spectrum
at large and small $\pi\eta$ invariant masses, respectively.

\begin{figure}[t]
\centerline{\includegraphics[width=0.85\textwidth]{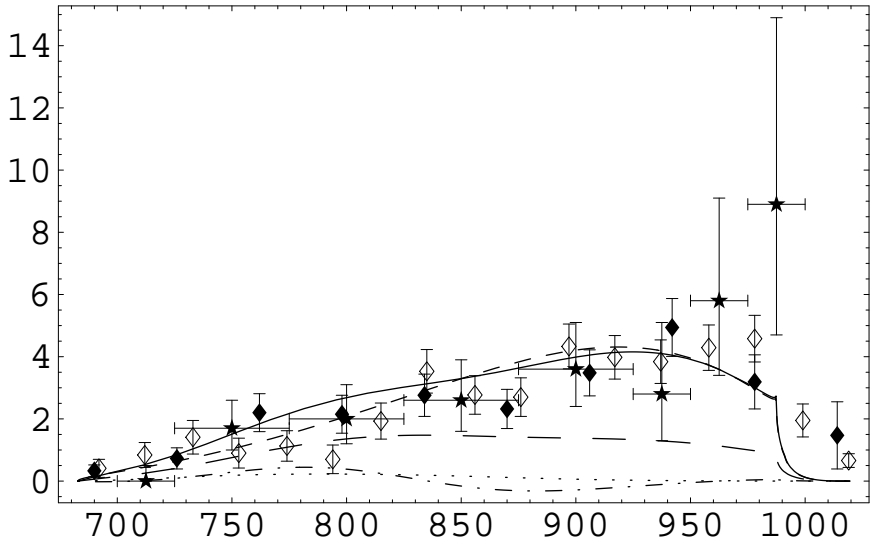}}  
\caption{\small
$dB(\phi\rightarrow\pi^0\eta\gamma)/dm_{\pi^0\eta}\times 10^7$ (in MeV$^{-1}$)   
\emph{versus} $m_{\pi^0\eta}$ (in MeV).
The curves follow the same conventions as in Fig.~\ref{dBdmphipi0pi0gamma}.
Experimental data are taken from Ref.~\cite{Achasov:2000ku} (solid star) and Ref.~\cite{Aloisio:2002bs}:
(open diamond) from $\eta\to\gamma\gamma$ and (solid diamond) from $\eta\to\pi^+\pi^-\pi^0$.}
\label{dBdmphipi0etagamma}
\end{figure}
The separate contributions from the L$\sigma$M, VMD and their interference,
as well as the total result, are displayed in Fig.~\ref{dBdmphipi0etagamma}.
We set $m_{a_0}=984.7$ MeV \cite{Eidelman:2004wy} and $\phi_P=41.8^\circ$ \cite{Aloisio:2002vm}.
As seen, the agreement with data is pretty good and the $a_0$ effects dominate.
The integrated branching ratio is
$B(\phi\to\pi^0\eta\gamma)_{\mbox{\scriptsize L$\sigma$M+VMD}}=8.2\times 10^{-5}$,
in accord with the experimental values in Refs.~\cite{Achasov:2000ku,Aloisio:2002bs}
and previous estimates \cite{Achasov:2001cj,Gokalp:2002xi,Palomar:2003rb,Achasov:2002ir}.
The sole contribution of the $a_0(980)$ amounts to
$B(\phi\to\pi^0\eta\gamma)_{\mbox{\scriptsize L$\sigma$M}}=7.8\times 10^{-5}$,
a value three times bigger than the chiral-loop prediction
$B(\phi\to\pi^0\eta\gamma)_{\chi}=2.9\times 10^{-5}$.
The contribution from vector exchange
---mainly $\rho$ exchange since the $\omega$ effects are calculated to be
less than 1\% of the total value---
is found to be
$B(\phi\to\pi^0\eta\gamma)_{\mbox{\scriptsize VMD}}=3.4\times 10^{-6}$,
a small effect, around 5\% of the final result including its interference with the scalar part,
which however should be taken into account in future analyses with more precise experimental data.
The obtained branching ratio is rather sensitive to the $\eta$-$\eta^\prime$ mixing angle.
If one uses $\theta_P=\arcsin(-1/3)\simeq -19.5^\circ$ ($\phi_P\simeq 35.3^\circ$)
the new result differs in a 10\%, 
a sizable effect which is however of the same order of the current experimental errors.
It is worth remarking that our predictions for the invariant mass distribution and branching ratio
of the $\phi\to\pi^0\eta\gamma$ decay are free from adjustable parameters,
\textit{i.e.}~the $a_0$ mass and the pseudoscalar mixing angle are fixed while
the $a_0$ couplings are predicted within the L$\sigma$M.
A main drawback of this approach is the large value of the $a_0\pi\eta$ coupling constant
which implies $\Gamma^{\mbox{\scriptsize L$\sigma$M}}_{a_0\to\pi\eta}=386$ MeV.
However, this large value is considerably reduced 
once the well-known Flatt\'e corrections for the $a_0$ propagator are introduced
\cite{Flatte:1976xu}.
The complete one-loop propagator used in this analysis is nothing else than
an improved version of the Flatt\'e formula taken into account the full energy dependence.

In summary, our present analysis supersedes Ref.~\cite{Bramon:2000vu}
incorporating not only the chiral constraints at low energy of the $\phi\to\pi^0\eta\gamma$ amplitude
but also the explicit dependence on the $\eta$-$\eta^\prime$ mixing angle
and the complete $a_0$ propagator in the corresponding four-pseudoscalar amplitude,
three new features which were not taken into account in the former work.

\subsection{$\rho\rightarrow\pi^0\pi^0\gamma$}
\label{rhopi0pi0gamma}
The last of the \emph{golden processes} is the $\rho\rightarrow\pi^0\pi^0\gamma$ decay,
where the $\sigma(600)$ contribution is supposed to dominate the scalar part of the amplitude
since the $f_0(980)$ effects are suppressed by kinematics.
In addition, there is also a well-known intermediate $\omega\pi$ contribution
which accounts for an important part of the total signal.
In spite of this sizable vector contribution,
the $\rho\rightarrow\pi^0\pi^0\gamma$ decay should be considered as an ideal process
for extracting the mass and decay width of the controversial and not well established $\sigma$ meson.
Therefore, a precise measurement of the branching ratio (already existing)
together with its invariant mass distribution (not yet available)
could help to fix unambiguously the $\sigma$ properties and
discriminate among different models predicting the size and shape of its contribution.
In this analysis, we provide a theoretical prediction for the branching ratio and mass spectrum
of the $\rho\rightarrow\pi^0\pi^0\gamma$ decay taking into account the contributions of the
$\sigma$ and $f_0$ scalar mesons, calculated within the framework of the L$\sigma$M,
as well as the $\omega\pi$ vector contribution given by the VMD model.

\begin{figure}[t]
\centerline{\includegraphics[width=0.85\textwidth]{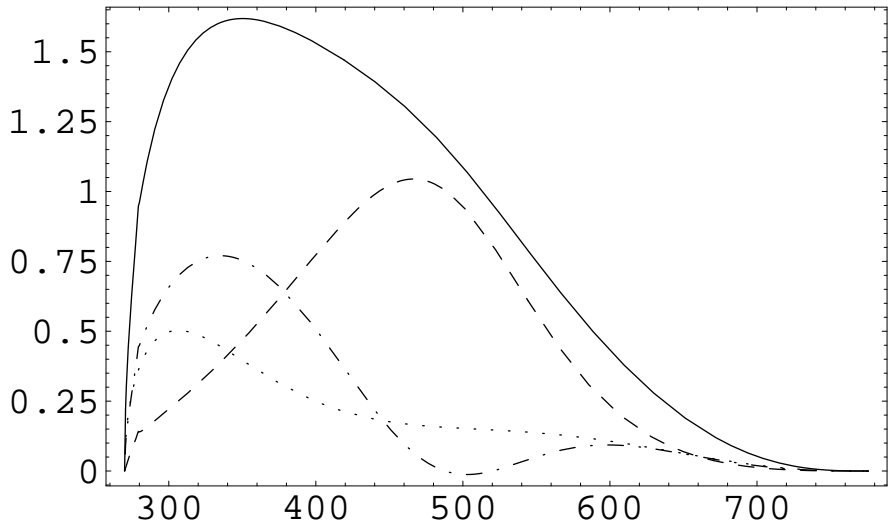}}  
\caption{\small
$dB(\rho\rightarrow\pi^0\pi^0\gamma)/dm_{\pi^0\pi^0}\times 10^7$ (in MeV$^{-1}$)   
\emph{versus} $m_{\pi^0\pi^0}$ (in MeV).
The dotted, dashed and dot-dashed lines correspond to the separate contributions 
from VMD, L$\sigma$M and their interference, respectively. 
The solid line is the total result.}
\label{dBdmrhopi0pi0gamma}
\end{figure}
\begin{figure}[t]
\centerline{\includegraphics[width=0.85\textwidth]{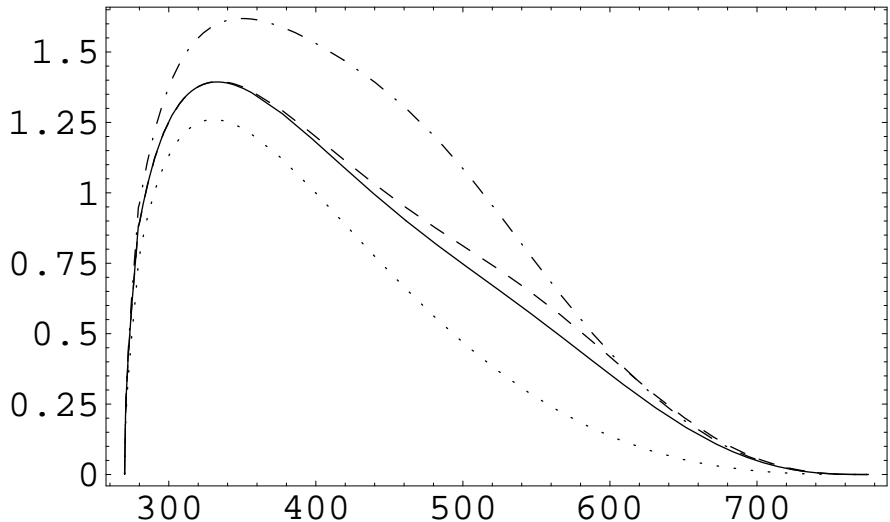}}  
\caption{\small 
$dB(\rho\rightarrow\pi^0\pi^0\gamma)/dm_{\pi^0\pi^0}\times 10^7\ (\mbox{MeV}^{-1})$ 
\emph{versus} $m_{\pi^0\pi^0}$ (MeV). 
The various predictions are for the input values:  
$m_\sigma=478$ MeV and $\Gamma_\sigma=324$ MeV    
\protect\cite{Aitala:2000xu} (solid line);   
$m_\sigma=478$ MeV \protect\cite{Aitala:2000xu} and
$\Gamma_\sigma=258$ MeV, as predicted by the L$\sigma$M (dot-dashed line); and  
$m_\sigma= 513$ MeV and $\Gamma_\sigma=335$ MeV  
\cite{Muramatsu:2002jp}, (dashed line).  
The chiral-loop prediction is also included for comparison (dotted line).} 
\label{dBdmvssigma}
\end{figure}

The separate contributions from the L$\sigma$M, VMD and their interference,
as well as the total result, are displayed in Fig.~\ref{dBdmrhopi0pi0gamma}.
We use $m_\sigma=478$ MeV \cite{Aitala:2000xu}, $\Gamma_\sigma=258$ MeV,
as required by the L$\sigma$M, $m_{f_0}=985$ MeV and $\phi_S=-8^\circ$.
The latter two values are fitted from the comparison to experimental data 
of our prediction for $\phi\rightarrow\pi^0\pi^0\gamma$
(see Sect.~\ref{phipi0pi0gamma} for details).
We have chosen a simple Breit-Wigner propagator for the $\sigma$ resonance.
This parametrization is considered in the literature as the preferred choice 
when dealing with broad scalar states.
As seen from Fig.~\ref{dBdmrhopi0pi0gamma},
the scalar contribution is peaked at $m_{\pi\pi}\simeq 500$ MeV,
as expected from a single resonance of that mass which is far from thresholds,
and the interference with the vector contribution is positive in the whole mass range
and suppressed in that particular region.
The integrated branching ratio obtained,
$B(\rho\to\pi^0\pi^0\gamma)_{\mbox{\scriptsize L$\sigma$M+VMD}}=4.2\times 10^{-5}$,
is in perfect agreement with the new experimental results
\cite{Achasov:2002jv,Akhmetshin:2003rg}
(confirming the older measurement \cite{Achasov:2000zr})
and previous analyses \cite{Gokalp:2003uf,Gokalp:2000ir,Marco:1999df,Palomar:2001vg,Oh:2003zz}.
The scalar part alone amounts to
$B(\rho\to\pi^0\pi^0\gamma)_{\mbox{\scriptsize L$\sigma$M}}=2.2\times 10^{-5}$,
which is clearly dominated by the $\sigma$ contribution
---$f_0$ effects are found to be less than 1\% of the total result---,
while the intermediate $\omega\pi^0$ exchange leads to
$B(\rho\to\pi^0\pi^0\gamma)_{\mbox{\scriptsize VMD}}=9.0\times 10^{-6}$.
Notice from the comparison of the former scalar contribution with the chiral-loop prediction,
$B(\rho\to\pi^0\pi^0\gamma)_{\chi}=1.1\times 10^{-5}$,
that taking into account the $\sigma$ effects in an explicit way is mandatory
in order to agree with experimental data.
If not, a lower less preferred value is obtained.
The $\sigma$ contribution is slightly sensitive to the scalar mixing angle
since $\phi_S\simeq -8^\circ$ and $g_{\sigma\pi\pi}\propto\cos\phi_S\simeq 1$
for small values of $\phi_S$.
However, it is quite dependent on the $\sigma$ properties, mass and width.
To study this dependence, we have redone our analysis applying different values for 
$m_\sigma$ and $\Gamma_\sigma$.
With the same mass value, $m_\sigma=478$ MeV,
but replacing the predicted L$\sigma$M width by $\Gamma_\sigma=324$ MeV \cite{Aitala:2000xu}
one gets $B(\rho\to\pi^0\pi^0\gamma)_{\mbox{\scriptsize L$\sigma$M}}=1.5\times 10^{-5}$,
which implies $B(\rho\to\pi^0\pi^0\gamma)_{\mbox{\scriptsize L$\sigma$M+VMD}}=3.3\times 10^{-5}$.
If instead, $m_\sigma=513$ MeV and $\Gamma_\sigma=335$ MeV
are taken from Ref.~\cite{Muramatsu:2002jp}, one gets 
$B(\rho\to\pi^0\pi^0\gamma)_{\mbox{\scriptsize L$\sigma$M}}=1.5\times 10^{-5}$
and $B(\rho\to\pi^0\pi^0\gamma)_{\mbox{\scriptsize L$\sigma$M+VMD}}=3.5\times 10^{-5}$.
Finally, the chiral-loop prediction, which is formally equivalent to the $m_\sigma\to\infty$ limit, gives
$B(\rho\rightarrow\pi^0\pi^0\gamma)_{\mbox{\scriptsize VMD}+\chi}=2.5\times 10^{-5}$.
The various predictions for the $\pi^0\pi^0$ mass spectrum using the former values of
$\sigma$ mass and width are shown in Fig.~\ref{dBdmvssigma}.
As seen, a detailed experimental check of this mass spectra
together with a more precise measurement of the branching ratio
would help to fix the mass and width of the elusive $\sigma$ meson.

This new analysis, taking into account the latest values of the Breit-Wigner mass and width of the
$\sigma$ resonance as well as the effects of the $f_0$ contribution for the first time,
allow us to update our former predictions for the shape and branching ratio of the
$\rho\rightarrow\pi^0\pi^0\gamma$ decay \cite{Bramon:2001un}
and confront them with the most recent experimental data.

\subsection{$\omega\rightarrow\pi^0\pi^0\gamma$}
This decay is less interesting from the point of view of extracting the properties of the $\sigma$ meson.
The reason is the following.
Because of $G$-parity pion loops are forbidden and
the chiral scalar contribution to the process is then driven by kaon loops.
Therefore, due to the large kaon mass, it is expected to be small as compared to the
$\rho\rightarrow\pi^0\pi^0\gamma$ case where pion loops are allowed.
In consequence, the $\omega\rightarrow\pi^0\pi^0\gamma$ transition is dominated by the
VMD contribution.
Recently, Guetta and Singer \cite{Guetta:2000ra} have explored the possibility of incorporating 
the pion loops through $\rho$-$\omega$ mixing effects.
However, as we see below, these effects are not enough to make the scalar contribution relevant.

Ignoring for the moment $\rho$-$\omega$ mixing,
\textit{i.e.}~assuming that the physical $\omega=\omega^{I=0}$ with no $I=1$ contaminations,
the well-known $\rho$ meson exchange contribution via the VMD decay chain
$\omega\to\rho\pi^0\to\pi^0\pi^0\gamma$ gives
$\Gamma(\omega\rightarrow\pi^0\pi^0\gamma)_{\mbox{\scriptsize VMD}}=244$ eV and
$B(\omega\rightarrow\pi^0\pi^0\gamma)_{\mbox{\scriptsize VMD}}=2.9\times 10^{-5}$,
a value which is more than two standard deviations below the experimental measurement in
Ref.~\cite{Achasov:2002jv}.
This result is improved using a momentum dependent width for the $\rho$ meson,
\begin{equation} 
\label{rhowidth} 
\Gamma_{\rho}(q^2)=\frac{m_\rho}{\sqrt{q^2}}\Gamma_{\rho} 
\left(\frac{q^2-4m^2_\pi}{m^2_\rho -4m^2_\pi}\right)^{3/2}\, 
\theta (q^2-4m^2_\pi)\ , 
\end{equation} 
leading to 
$\Gamma(\omega\rightarrow\pi^0\pi^0\gamma)_{\rm VMD}=272$ eV,
which is some 12\% larger than the previous value, as already noticed in Ref.~\cite{Guetta:2000ra}.
There is also another contribution to the $\omega\rightarrow\pi^0\pi^0\gamma$ amplitude 
coming from chiral loops. 
However, as stated before, this chiral loop contribution  
(given only by kaon loops in the good isospin limit with $\omega =\omega^{I=0}$) 
is very small, 
$\Gamma(\omega\rightarrow\pi^0\pi^0\gamma)_{\chi}=1.3$ eV,
and can be safely neglected.   
Its improved version taking into account  $\sigma$ resonance effects is also negligible,
$\Gamma(\omega\to\pi^0\pi^0\gamma)_{\mbox{\scriptsize L$\sigma$M}}=1.5$ eV,
due to the suppression of the $g_{\sigma K\bar K}$ coupling for $m_\sigma\simeq m_K$
(see Sec.~\ref{phipi0pi0gamma} for details). 

\begin{figure}[t] 
\centerline{\includegraphics[width=0.85\textwidth]{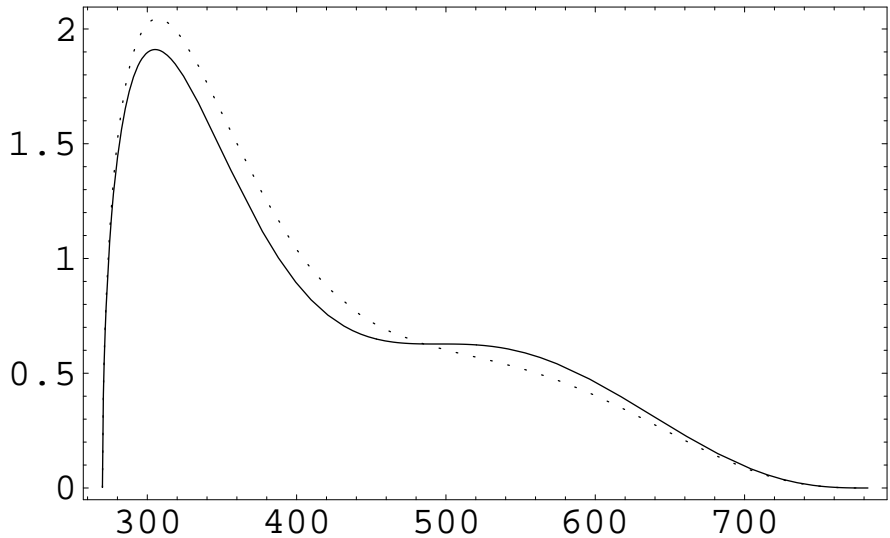}}  
\caption{\small 
$dB(\omega\rightarrow\pi^0\pi^0\gamma)/dm_{\pi^0\pi^0}\times 10^7\ (\mbox{MeV}^{-1})$ 
\emph{versus} $m_{\pi^0\pi^0}$ (MeV). 
The plotted curves are for $m_\sigma=478$ MeV and $\Gamma_\sigma=324$ MeV (solid line)
and dropping all $\sigma$ meson effects  ---chiral loop prediction--- (dotted line).} 
\label{dBdmomegapi0pi0gamma}
\end{figure}

Taking now into account $\rho$-$\omega$ mixing, there is an indirect contribution to the
$\omega\rightarrow\pi^0\pi^0\gamma$ amplitude that is written as
${\cal A}^{I=0}(\omega\rightarrow\pi^0\pi^0\gamma)_{\rm VMD}+
   \epsilon {\cal A}(\rho\rightarrow\pi^0\pi^0\gamma)$,
where
${\cal A}^{I=0}(\omega\rightarrow\pi^0\pi^0\gamma)_{\rm VMD}=
   \frac{1}{3} {\cal A}(\rho\rightarrow\pi^0\pi^0\gamma)_{\rm VMD}$
---the proportionality factor follows from the $SU(3)$-symmetric Lagrangian (\ref{VMDLag})---
and $\epsilon$ is the $\rho$-$\omega$ mixing parameter given by\footnote{
A further effect of this $\rho$-$\omega$ mixing is to replace the $\rho$ propagator in
${\cal A}^{I=0}$ by 
\begin{equation} 
\label{Drho} 
\frac{1}{D_\rho(s)}\rightarrow\frac{1}{D^\epsilon_\rho(s)}=\frac{1}{D_\rho(s)} 
\left(1+\frac{g_{\omega\pi\gamma}}{g_{\rho\pi\gamma}} 
        \frac{{\cal M}^2_{\rho\omega}}{D_\omega(s)}\right)\ , 
\end{equation} 
with $D_V(s)=s-m^2_V+i\,m_V\Gamma_V$ for $V=\rho, \omega$ and 
$g_{\omega\pi\gamma}/g_{\rho\pi\gamma}=3$ in the $SU(3)$-symmetric VMD framework.} 
\begin{equation} 
\label{epsilon} 
\epsilon \equiv \frac{{\cal M}^2_{\rho\omega}} 
              {m^2_{\omega}-m^2_{\rho}-i(m_\omega\Gamma_{\omega}-m_\rho\Gamma_{\rho})} 
\simeq -0.003+i\,0.034\ , 
\end{equation} 
with  
${\cal M}^2_{\rho\omega}(m^2_\rho)=(-3800\pm 370)\ \mbox{MeV}^2$ \cite{O'Connell:1995wf}. 
Approximating the new, isospin violating term of $\omega\rightarrow\pi^0\pi^0\gamma$
by the VMD contribution $\epsilon {\cal A}(\rho\rightarrow\pi^0\pi^0\gamma)_{\rm VMD}$,
one increases the previous estimate to   
$\Gamma(\omega\rightarrow\pi^0\pi^0\gamma)=296$ eV.
However, a more complete treatment, with  
${\cal A}(\omega\rightarrow\pi^0\pi^0\gamma)=  
 {\cal A}^{I=0}(\omega\rightarrow\pi^0\pi^0\gamma)_{\rm VMD}+\epsilon 
 {\cal A}(\rho\rightarrow\pi^0\pi^0\gamma)_{\mbox{\scriptsize VMD+L$\sigma$M}}$, 
seems preferable. 
The $\pi^0\pi^0$ invariant mass spectra corresponding to this amplitude  
have been calculated with the same $m_\sigma$ and $\Gamma_\sigma$ values
introduced for the $\rho\rightarrow\pi^0\pi^0\gamma$ case.  
The sensitivity on the $\sigma$ parameters is seen to be minimal and all the results 
almost coincide with the curve for $m_\sigma=478$ MeV and $\Gamma_\sigma=324$ MeV 
\cite{Aitala:2000xu} plotted in Fig.~\ref{dBdmomegapi0pi0gamma}. 

The integrated width and branching ratio for the former values are predicted to be  
$\Gamma(\omega\rightarrow\pi^0\pi^0\gamma)_{\mbox{\scriptsize VMD+L$\sigma$M}}=297$ eV and  
$B(\omega\rightarrow\pi^0\pi^0\gamma)_{\mbox{\scriptsize VMD+L$\sigma$M}}=3.5\times 10^{-5}$.
If all $\sigma$ meson effects are neglected ---chiral-loop prediction---  one then obtains
$\Gamma(\omega\rightarrow\pi^0\pi^0\gamma)_{\mbox{\scriptsize VMD+$\chi$}}=307$ eV and  
$B(\omega\rightarrow\pi^0\pi^0\gamma)_{\mbox{\scriptsize VMD+$\chi$}}=3.6\times 10^{-5}$,
only a 4\% above the previous results and hardly distinguishable
(see Fig.~\ref{dBdmomegapi0pi0gamma}).
Notice that our values are still substantially lower than the central value reported in
Ref.~\cite{Achasov:2002jv}.
The reason is that we are using an $SU(3)$-symmetric formalism where
all the VVP and VP$\gamma$ couplings are deduced from the single VPP coupling $g$
taken from $\Gamma(\rho\to\pi^+\pi^-)$.
In this formalism, the product of couplings
$g_{\omega\rho\pi}g_{\rho^0\pi^0\gamma}=\frac{G^2 e}{3\sqrt{2}g}=3.68$ GeV$^{-2}$
for $|g|=4.24$ and $g_{\omega\rho\pi}\equiv G=14.8$  GeV$^{-1}$.
Alternatively, one could use the experimental values
$g_{\omega\rho\pi}^{\rm exp}=16.7$  GeV$^{-1}$
and $g_{\rho^0\pi^0\gamma}^{\rm exp}=0.25$ GeV$^{-1}$,
extracted from the study of the process $e^+e^-\to\omega\pi^0\to\pi^0\pi^0\gamma$ above 1 GeV
at CMD-2 \cite{Akhmetshin:2003ag} and the current value of
$\Gamma(\rho^0\to\pi^0\gamma)=(91\pm 13)$ keV \cite{Eidelman:2004wy},
to fix
$g_{\omega\rho\pi}^{\rm exp}g_{\rho^0\pi^0\gamma}^{\rm exp}=4.24$ GeV$^{-2}$,
which is 15\% larger than the $SU(3)$-symmetric result.
This enhancement applied to the squared amplitude gives
$\Gamma(\omega\rightarrow\pi^0\pi^0\gamma)_{\mbox{\scriptsize VMD+L$\sigma$M}}=394$ eV and  
$B(\omega\rightarrow\pi^0\pi^0\gamma)_{\mbox{\scriptsize VMD+L$\sigma$M}}=4.6\times 10^{-5}$,
a value which is still below the experimental result in Ref.~\cite{Achasov:2002jv}
by more than one standard deviation.
Our prediction is in fair agreement with a previous analysis in Ref.~\cite{Palomar:2001vg}
but clearly disagrees with Ref.~\cite{Gokalp:2003dr}.

In conclusion, our final results for the branching ratio of the
$\omega\rightarrow\pi^0\pi^0\gamma$ decay is
$B(\omega\rightarrow\pi^0\pi^0\gamma)_{\mbox{\scriptsize VMD+L$\sigma$M}}=$
(3.5--4.6)$\times 10^{-5}$, where the numeric interval takes into account the departure from the
$SU(3)$-symmetric VMD prediction.
Therefore, $4.6\times 10^{-5}$ seems to be the maximum branching ratio acceptable within
VMD model and additional theoretical study would be required to explain the large value of
$B(\omega\rightarrow\pi^0\pi^0\gamma)$.

\subsection{$\rho\rightarrow\pi^0\eta\gamma$ and $\omega\rightarrow\pi^0\eta\gamma$}
The scalar contribution to these two processes is driven through a loop of charged kaons
since the $I=1$ of the final state $\pi\eta$ forbids a loop of pions due to Bose symmetry.
Therefore, scalar effects are expected to be suppressed,
 because of the large kaon mass,
and vector meson exchange contributions should dominate.
Consequently, these decays seem to be not very promising from the point of view of 
extracting the $a_0(980)$ properties.
However, they can still be used to test the hypothesis of a dominant intermediate vector contribution
as soon as experimental data is available.
Unfortunately, the size of the predicted branching ratios,
of the order of $10^{-9}$ for $\rho\rightarrow\pi^0\eta\gamma$ and
$10^{-7}$ for $\omega\rightarrow\pi^0\eta\gamma$,
make of their experimental analyses a very difficult task not feasible in the near future.
At the moment, only the upper limit
$B(\omega\rightarrow\pi^0\eta\gamma)<3.3\times 10^{-5}$ at 90\% CL
has been obtained \cite{Akhmetshin:2003rg}.

\begin{figure}[t]
\centerline{\includegraphics[width=0.5\textwidth]{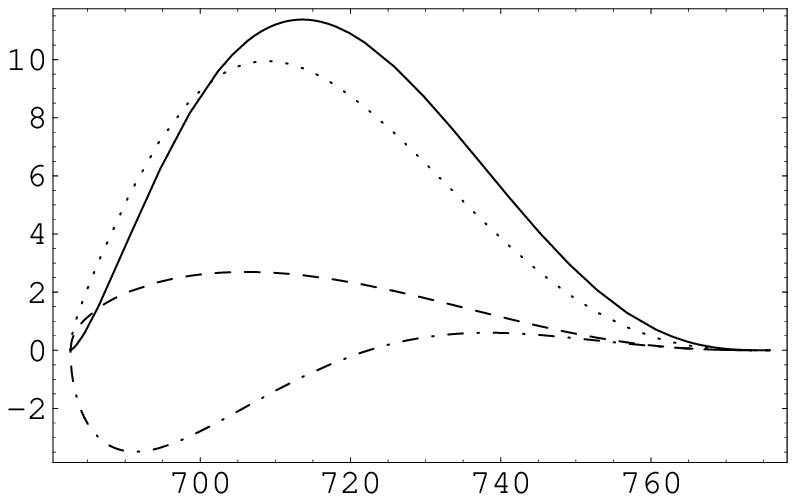}
                    \includegraphics[width=0.5\textwidth]{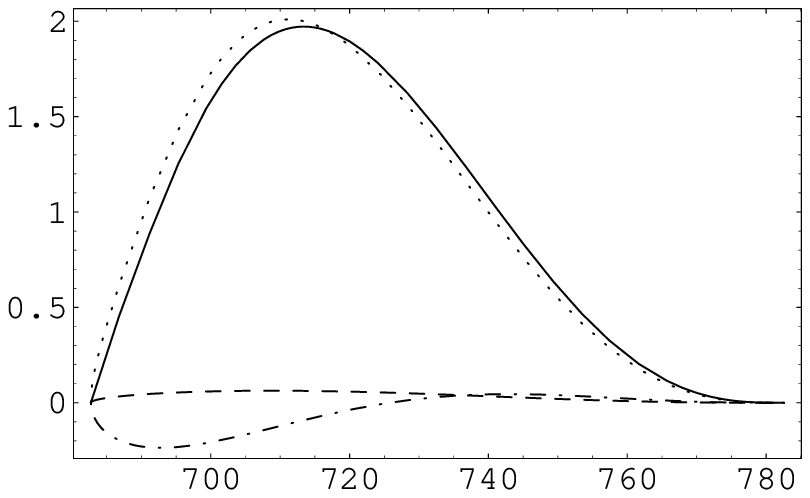}}  
\caption{\small
$dB(\rho\rightarrow\pi^0\eta\gamma)/dm_{\pi^0\eta}\times 10^{12}$ (MeV$^{-1}$)
---\emph{left plot}--- and
$dB(\omega\rightarrow\pi^0\eta\gamma)/dm_{\pi^0\eta}\times 10^{9}$ (MeV$^{-1}$)
---\emph{right plot}---   
\emph{versus} $m_{\pi^0\eta}$ (MeV).
The curves follow the same conventions as in Fig.~\ref{dBdmrhopi0pi0gamma}.}
\label{dBdmrhoomegapi0etagamma}
\end{figure}
\begin{figure}[t]
\centerline{\includegraphics[width=0.5\textwidth]{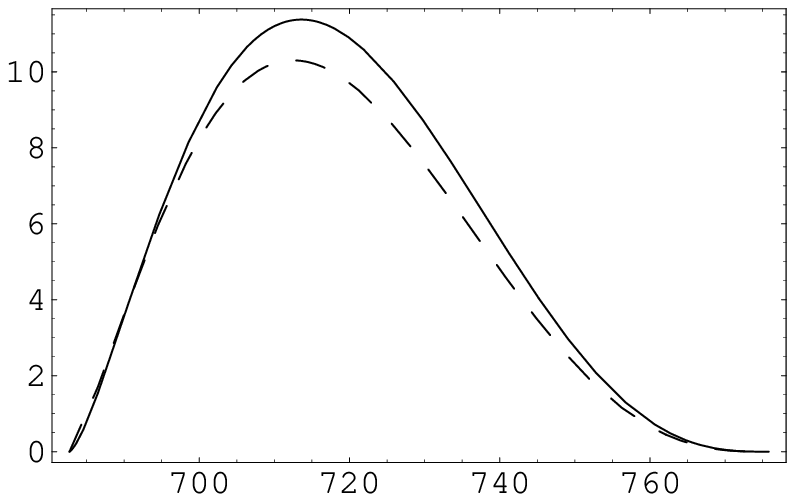}
                    \includegraphics[width=0.5\textwidth]{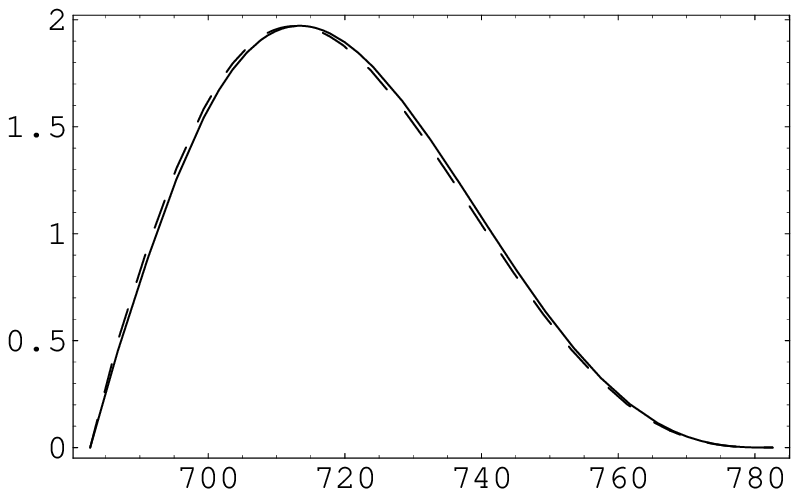}}  
\caption{\small
$dB(\rho\rightarrow\pi^0\eta\gamma)/dm_{\pi^0\eta}\times 10^{12}$ (MeV$^{-1}$)
---\emph{left plot}--- and
$dB(\omega\rightarrow\pi^0\eta\gamma)/dm_{\pi^0\eta}\times 10^{9}$ (MeV$^{-1}$)
---\emph{right plot}---   
\emph{versus} $m_{\pi^0\eta}$ (MeV).
The plotted curves are for an explicit $a_0$ resonance with $m_{a_0}=984.7$ MeV (solid line)
and dropping all $a_0$ meson effects  ---chiral loop prediction--- (dashed line).} 
\label{dBdmrhoomegaLsMvsChPT}
\end{figure}

The contributions to $\rho\rightarrow\pi^0\eta\gamma$ and $\omega\rightarrow\pi^0\eta\gamma$
from the L$\sigma$M, VMD and their interference, as well as the total result, are shown
on the left and right hand sides of Fig.~\ref{dBdmrhoomegapi0etagamma}, respectively.
We have used the same input values for the $a_0$ mass and pseudoscalar mixing angle
as in the $\phi\rightarrow\pi^0\eta\gamma$ analysis.
As seen, the vector exchange contribution is dominant in both cases and the integrated branching ratio
gives $B(\rho\rightarrow\pi^0\eta\gamma)_{\mbox{\scriptsize VMD}}=4.5\times 10^{-10}$ and
$B(\omega\rightarrow\pi^0\eta\gamma)_{\mbox{\scriptsize VMD}}=9.9\times 10^{-8}$.
The difference of two orders of magnitude in branching ratio is easily understood
in terms of the ratio
$B_{\rho\rightarrow\pi^0\eta\gamma}^{\rm VMD}/B_{\omega\rightarrow\pi^0\eta\gamma}^{\rm VMD}
\sim\Gamma_\omega/(9\Gamma_\rho)$
where the numerical factor is deduced from Eq.~(\ref{pi0etagammacoupling}).
Although smaller in size, the scalar contribution to the $\rho\rightarrow\pi^0\eta\gamma$ decay
is more relevant, as compared to the vector one, than for the $\omega\rightarrow\pi^0\eta\gamma$ case.
In the former, its integrated branching ratio amounts to
$B(\rho\rightarrow\pi^0\eta\gamma)_{\mbox{\scriptsize L$\sigma$M}}=1.3\times 10^{-10}$,
around 25\% of the total signal,
while in the latter
$B(\omega\rightarrow\pi^0\eta\gamma)_{\mbox{\scriptsize L$\sigma$M}}=3.4\times 10^{-9}$,
less than 5\% of the signal.
If the $a_0$ resonance is removed from the analysis, the prediction of chiral loops is
$B(\rho\rightarrow\pi^0\eta\gamma)_{\chi}=6.3\times 10^{-11}$ and
$B(\omega\rightarrow\pi^0\eta\gamma)_{\chi}=1.6\times 10^{-9}$,
so in both cases a factor of two of enhancement is achieved when the $a_0$ is introduced explicitly.
In view of all the above discussion, it would be preferable as long as experimentally possible
to look for the $a_0$ effects in the $\rho\rightarrow\pi^0\eta\gamma$ decay
since in the $\omega\rightarrow\pi^0\eta\gamma$ they are not discernible
(see Fig.~\ref{dBdmrhoomegaLsMvsChPT} for comparison).
However, as stated before, $\rho\rightarrow\pi^0\eta\gamma$ should be considered as a
very rare decay and hence very hard to detect.
Fortunately, one should keep in mind that there is a \emph{golden process}
for extracting the $a_0$ properties, which is obviously the $\phi\rightarrow\pi^0\eta\gamma$ decay
discussed previously.
Finally, our final results including scalar and vector exchange contributions are
$B(\rho\rightarrow\pi^0\eta\gamma)_{\mbox{\scriptsize VMD+L$\sigma$M}}=5.2\times 10^{-10}$ and
$B(\omega\rightarrow\pi^0\eta\gamma)_{\mbox{\scriptsize VMD+L$\sigma$M}}=9.7\times 10^{-8}$,
which should be compared with
$B(\rho\rightarrow\pi^0\eta\gamma)_{\mbox{\scriptsize VMD+$\chi$}}=4.7\times 10^{-10}$ and
$B(\omega\rightarrow\pi^0\eta\gamma)_{\mbox{\scriptsize VMD+$\chi$}}=9.7\times 10^{-8}$
in absence of scalar effects.

\section{Conclusions}
\label{Conclusions}
In this work we have performed an extensive and exhaustive analysis of
the scalar and vector meson exchange contributions to the $V\to P^0P^0\gamma$ decays
with $V=\rho, \omega, \phi$ and $P^0P^0=\pi^0\pi^0, \pi^0\eta, K^0\bar K^0$.
The scalar contributions, which turn out to be dominant in most of the interesting cases,
have been studied within the framework of the L$\sigma$M.
This model has the advantage of incorporating not only the scalar resonances in an explicit way
but also the constraints required by chiral symmetry.
The complementarity between ChPT and the L$\sigma$M have been used to parametrize
the needed scalar amplitudes making explicitly the effect of scalar meson poles
while keeping the correct behaviour at low invariant masses.
Therefore, our treatment of the scalar contributions can be considered as an improvement
of the corresponding chiral-loop predictions.
Moreover, we have updated the vector exchange contributions.
A summary of the results obtained is presented in Table \ref{tableresults}.
Our theoretical predictions are all compatible with experimental data.
This nice agreement is achieved not only for the branching ratios but also for the available mass spectra.
In general, we have shown that
$\phi\rightarrow\pi^0\pi^0\gamma$, $\phi\rightarrow\pi^0\eta\gamma$ and
$\rho\rightarrow\pi^0\pi^0\gamma$
can be used to extract relevant information on the properties of the
$f_0(980)$, $a_0(980)$ and $\sigma(600)$ scalar mesons, respectively,
while $\omega\rightarrow\pi^0\pi^0\gamma$ and $(\rho, \omega)\rightarrow\pi^0\eta\gamma$
cannot be used due to the smallness of the scalar contribution.
$\phi\rightarrow K^0\bar K^0\gamma$ could also be used in the near future
for extracting the properties of the $f_0$ and $a_0$ resonances.
\begin{table}
    \centerline{
    \begin{tabular}{cccc}
	\hline\hline\\[-2ex]
	$B(V\rightarrow P^0P^0\gamma)$        & L$\sigma$M   & VMD  & L$\sigma$M+VMD\\[0.5ex]
	\hline\hline\\[-2ex]
	$\phi\rightarrow\pi^0\pi^0\gamma$       & $1.00\times 10^{-4}$ & $8.3\times 10^{-6}$
								     &$1.11\times 10^{-4}$\\[0.5ex]
	$\phi\rightarrow\pi^0\eta\gamma$         & $7.8\times 10^{-5}$    & $3.4\times 10^{-6}$
								     &$8.2\times 10^{-5}$\\[0.5ex]
	$\phi\rightarrow K^0\bar K^0\gamma$  & $7.5\times 10^{-8}$    & $2.0\times 10^{-12}$
								      &$7.5\times 10^{-8}$\\[0.5ex]
	\hline\\[-2ex]
	$\rho\rightarrow\pi^0\pi^0\gamma$       & $2.2\times 10^{-5}$    & $9.0\times 10^{-6}$
								     & $4.2\times 10^{-5}$\\[0.5ex]
	$\rho\rightarrow\pi^0\eta\gamma$         & $1.3\times 10^{-10}$  & $4.5\times 10^{-10}$
								     & $5.2\times 10^{-10}$\\[0.5ex]
	\hline\\[-2ex]
	$\omega\rightarrow\pi^0\pi^0\gamma$ & $1.7\times 10^{-7}$   & (2.9--3.5)$\times 10^{-5}$
								      & (3.5--4.6)$\times 10^{-5}$\\[0.5ex]
	$\omega\rightarrow\pi^0\eta\gamma$   & $3.4\times 10^{-9}$   & $9.9\times 10^{-8}$
								      & $9.7\times 10^{-8}$\\[0.5ex]
	\hline\hline
    \end{tabular}
    }
    \caption{Scalar (L$\sigma$M) and vector meson (VMD) exchange contributions to
                    $V\rightarrow P^0P^0\gamma$ decays.}
    \label{tableresults}
\end{table}
The following comments are for each process in particular.
For the $\phi\rightarrow\pi^0\pi^0\gamma$ decay, we have seen that the $f_0$ exchange dominates
whereas the contributions of the $\sigma$ and intermediate vectors are of 10\% each.
The dependence of our prediction on the $f_0$ mass and the scalar mixing angle
could be used to obtain relevant information on these two parameters.
We have also provided an explanation for the suppression of the $\sigma$ contribution.
For $\phi\rightarrow\pi^0\eta\gamma$, we have a parameter-free prediction
which is dominated by the $a_0$ exchange and where the vector exchange is only of 5\%.
For the not yet measured $\phi\rightarrow K^0\bar K^0\gamma$ decay,
we have shown that the $f_0$ contributes more strongly than the $a_0$
while other contributions are negligible.
Our prediction seems to be very sensitive to the scalar mixing angle.
A measurement of this process at DAFNE-2 would be welcome and can serve
as an additional test of the whole approach.
It is also confirmed that this process does not pose a background problem for testing
CP-violation at DA$\Phi$NE.
Concerning the $\rho\rightarrow\pi^0\pi^0\gamma$ decay,
our analysis predicts an important contribution from the $\sigma$ resonance
which accounts for around half of the total signal.
A comparison with experimental data seems to prefer a low mass and moderately narrow
$\sigma(600)$.
Further experimental improvement on this decay  is required before drawing definite conclusions.
On the contrary, the scalar contribution to the $\omega\rightarrow\pi^0\pi^0\gamma$ decay
is negligible and the process is dominated by the intermediate $\rho$ meson exchange.
However, this vector contribution complemented with $\rho$-$\omega$ mixing effects and
the use of a momentum dependent $\rho$ width
is not enough to reach the large experimental branching ratio which remains unexplained.
Finally, the decays $(\rho, \omega)\rightarrow\pi^0\eta\gamma$ are not very appealing
since the predicted size of their branching ratios is far from being observed in the near future.
In conclusion, higher accuracy data and more refined theoretical analyses
would contribute decisively to clarify the sector of the lowest lying scalar states.
In this line, our work aims to be one step forward.

\section*{Acknowledgements}
This work was supported in part by the Ramon y Cajal program,
the Ministerio de Educaci\'on y Ciencia under grant FPA2005-02211,
the EU Contract No.~MRTN-CT-2006-035482, ``FLAVIAnet'', and
the Generalitat de Catalunya under grant 2005-SGR-00994.

\appendix
\section{Appendix}
\label{appendix}
\subsection{Scalar amplitudes}
\label{scalaramp}
\subsubsection*{\boldmath $K^+K^-\to K^0\bar K^0$}
The $K^+K^-\to K^0\bar K^0$ amplitude in the L$\sigma$M is
\begin{eqnarray}
{\cal A}_{K^+K^-\rightarrow K^0\bar K^0}^{\mbox{\scriptsize L$\sigma$M}}
&=& g_{K^+ K^- K^0\bar K^0}
    -\frac{g_{\sigma K^+K^-}g_{\sigma K^0\bar K^0}}{s-m^2_\sigma}
    -\frac{g_{f_0 K^+K^-}g_{f_0 K^0\bar K^0}}{s-m^2_{f_0}}\nonumber\\
&&  -\frac{g_{a_0 K^+K^-}g_{a_0 K^0\bar K^0}}{s-m^2_{a_0}}
    -\frac{g^2_{a_{0}^\pm K^\mp K^{0(\bar 0)}}}{t-m^2_{a_{0}}}\ ,
\end{eqnarray}
where the $SPP$ couplings in the isospin limit
(see Refs.~\cite{Napsuciale:1998ip,Tornqvist:1999tn} for details)
are
\begin{equation}
\label{gKpKmK0K0bar}
\begin{array}{c}
g_{\sigma K^+K^-}=g_{\sigma K^0\bar K^0}\equiv g_{\sigma K\bar K}=
\frac{m^2_K-m^2_\sigma}{2f_K}(\cos\phi_S-\sqrt{2}\sin\phi_S)\ ,\\[2ex]
g_{f_0 K^+K^-}=g_{f_0 K^0\bar K^0}\equiv g_{f_0 K\bar K}=
\frac{m^2_K-m^2_{f_0}}{2f_K}(\sin\phi_S+\sqrt{2}\cos\phi_S)\ ,\\[2ex]
g_{a_0 K^+K^-}=-g_{a_0 K^0\bar K^0}=\frac{g_{a_0^\pm K^\mp K^{0(\bar 0)}}}{\sqrt{2}}=
\frac{m^2_K-m^2_{a_0}}{2f_K}\ ,
\end{array}
\end{equation}
and $\phi_S$ is the scalar mixing angle in the quark-flavour basis defined as
\begin{equation}
\sigma=\cos\phi_S\sigma_q-\sin\phi_S\sigma_s\ ,
\qquad
f_0=\sin\phi_S\sigma_q+\cos\phi_S\sigma_s\ ,
\end{equation}
with $\sigma_q\equiv\frac{1}{\sqrt{2}}(u\bar u+d\bar d)$ and $\sigma_s\equiv s\bar s$.
The $g_{K^+ K^- K^0\bar K^0}$ coupling is fixed by imposing that the
${\cal A}_{K^+K^-\rightarrow K^0\bar K^0}^{\mbox{\scriptsize L$\sigma$M}}$
vanishes in the soft-kaon limit (either $p\to 0$ or $p^\prime\to 0$) implying $s=t=u=m^2_K$.
The resulting expression,
\begin{equation}
g_{K^+ K^- K^0\bar K^0}=
\frac{g^2_{\sigma K\bar K}}{m^2_K-m^2_\sigma}+\frac{g^2_{f_0 K\bar K}}{m^2_K-m^2_{f_0}}+
\frac{g_{a_0 K^+K^-}g_{a_0 K^0\bar K^0}}{m^2_K-m^2_{a_0}}+
\frac{g^2_{a_{0}^\pm K^\mp K^{0(\bar 0)}}}{m^2_K-m^2_{a_{0}}}\ ,
\end{equation}
has been explicitly checked within the L$\sigma$M.
In consequence,
\begin{eqnarray}
{\cal A}_{K^+K^-\rightarrow K^0\bar K^0}^{\mbox{\scriptsize L$\sigma$M}}
&=&
\frac{g^2_{\sigma K\bar K}}{m^2_K-m^2_\sigma}\frac{s-m^2_K}{s-m^2_\sigma}+
\frac{g^2_{f_0 K\bar K}}{m^2_K-m^2_{f_0}}\frac{s-m^2_K}{s-m^2_{f_0}}\nonumber\\
&+&
\frac{g_{a_0 K^+K^-}g_{a_0 K^0\bar K^0}}{m^2_K-m^2_{a_0}}\frac{s-m^2_K}{s-m^2_{a_0}}+
\frac{g^2_{a_{0}^\pm K^\mp K^{0(\bar 0)}}}{m^2_K-m^2_{a_{0}}}\frac{t-m^2_K}{t-m^2_{a_0}}\ ,
\end{eqnarray}
and utilizing the definite forms of the couplings in Eq.~(\ref{gKpKmK0K0bar}), one obtains
\begin{eqnarray}
\label{AKpKmK0K0barLsMt}
{\cal A}_{K^+K^-\rightarrow K^0\bar K^0}^{\mbox{\scriptsize L$\sigma$M}}
&=&
\frac{s-m^2_{K}}{4f_K^2}
\left[
\frac{m^2_K-m^2_{\sigma}}{s-m^2_\sigma}({\rm c}\phi_S -\sqrt{2}{\rm s}\phi_S)^2\right.\nonumber\\
&& \left.
+\frac{m^2_K-m^2_{f_0}}{s-m^2_{f_0}}({\rm s}\phi_S +\sqrt{2}{\rm c}\phi_S)^2
-\frac{m^2_K-m^2_{a_0}}{s-m^2_{a_0}}\right]\nonumber\\
&&
+\frac{t-m^2_K}{2f_K^2}\frac{m^2_K-m^2_{a_{0}}}{t-m^2_{a_0}}\ .
\end{eqnarray}
In order to implement this amplitude into
${\cal A}_{\phi\to K^0 \bar K^0\gamma}^{\mbox{\scriptsize L$\sigma$M}}$
preserving the chiral-loop prediction for infinite scalar masses,
the former has to become first only $s$-dependent.
As explained in Sect.~\ref{sectLsM},
this is achieved replacing the $t$-dependent part of Eq.~(\ref{AKpKmK0K0barLsMt})
by the result of subtracting from the chiral-loop amplitude (\ref{AKpKmK0K0barChPT})
the infinite mass limit of the $s$-channel scalar contributions.
In doing so, one finally gets the scalar amplitude (\ref{AKpKmK0K0barChPTLsM})
where the naive scalar poles have been already substituted by more appropriate scalar propagators,
namely a Breit-Wigner for the $\sigma$
and complete one-loop propagators for the $f_0$ and $a_0$ (see Sect.~\ref{propagators}).
The amplitude (\ref{AKpKmK0K0barChPTLsM}), now only $s$-dependent, is then incorporated in
${\cal A}_{\phi\to K^0 \bar K^0\gamma}^{\mbox{\scriptsize L$\sigma$M}}$
to give the result shown in Eq.~(\ref{AphiK0K0barLsM}).

\subsubsection*{\boldmath $K^+K^-\to\pi^0\pi^0$}
The $K^+K^-\to\pi^0\pi^0$ amplitude in the L$\sigma$M is
\begin{eqnarray}
{\cal A}_{K^+K^-\rightarrow\pi^0\pi^0}^{\mbox{\scriptsize L$\sigma$M}}
&=& g_{K^+ K^-\pi^0\pi^0}
    -\frac{g_{\sigma K^+K^-}g_{\sigma\pi^0\pi^0}}{s-m^2_\sigma}
    -\frac{g_{f_0 K^+K^-}g_{f_0\pi^0\pi^0}}{s-m^2_{f_0}}\nonumber\\
&&  -g^2_{\kappa^\mp K^\pm\pi^0}
         \left(\frac{1}{t-m^2_\kappa}+\frac{1}{u-m^2_\kappa}\right)\ ,
\end{eqnarray}
where the required $SPP$ couplings, besides those of Eq.~(\ref{gKpKmK0K0bar}), are
\begin{equation}
\label{gKpKmpi0pi0}
\begin{array}{c}
g_{\sigma\pi^0\pi^0}=\frac{m^2_\pi-m^2_\sigma}{f_\pi}\cos\phi_S\ ,
\qquad
g_{f_0\pi^0\pi^0}=\frac{m^2_\pi-m^2_{f_0}}{f_\pi}\sin\phi_S\ ,\\[2ex]
g_{\kappa^\mp K^\pm\pi^0}=\frac{m^2_\pi-m^2_\kappa}{2f_K}=\frac{m^2_K-m^2_\kappa}{2f_\pi}\ .
\end{array}
\end{equation}
Notice that $g_{\kappa^\mp K^\pm\pi^0}$ can be written in two different equivalent forms.
Using the soft-pion limit ($p_\pi\to 0$), which implies $s=m^2_\pi$ and $t=u=m^2_K$,
the coupling $g_{K^+ K^-\pi^0\pi^0}$ is fixed to
\begin{equation}
g_{K^+ K^-\pi^0\pi^0}=
\frac{g_{\sigma K^+K^-}g_{\sigma\pi^0\pi^0}}{m^2_\pi-m^2_\sigma}+
\frac{g_{f_0 K^+K^-}g_{f_0\pi^0\pi^0}}{m^2_\pi-m^2_{f_0}}+
2\frac{g^2_{\kappa^\mp K^\pm\pi^0}}{m^2_K-m^2_\kappa}\ ,
\end{equation}
and the amplitude to
\begin{eqnarray}
{\cal A}_{K^+K^-\rightarrow\pi^0\pi^0}^{\mbox{\scriptsize L$\sigma$M}}
&=& \frac{g_{\sigma K^+K^-}g_{\sigma\pi^0\pi^0}}{m^2_\pi-m^2_\sigma}
         \frac{s-m^2_\pi}{s-m^2_\sigma}+
         \frac{g_{f_0 K^+K^-}g_{f_0\pi^0\pi^0}}{m^2_\pi-m^2_{f_0}}
         \frac{s-m^2_\pi}{s-m^2_{f_0}}\nonumber\\
&& +\frac{g^2_{\kappa^\mp K^\pm\pi^0}}{m^2_K-m^2_\kappa}
         \left(\frac{t-m^2_K}{t-m^2_\kappa}+ \frac{u-m^2_K}{u-m^2_\kappa}\right)\ ,
\end{eqnarray}
that after utilizing the explicit form of the couplings in
Eqs.~(\ref{gKpKmK0K0bar}) and (\ref{gKpKmpi0pi0}) becomes
\begin{eqnarray}
\label{AKpKmpi0pi0LsMtu}
{\cal A}_{K^+K^-\rightarrow\pi^0\pi^0}^{\mbox{\scriptsize L$\sigma$M}}
&=& \frac{s-m^2_\pi}{2f_\pi f_K}
\left[\frac{m^2_K-m^2_\sigma}{s-m^2_\sigma}\cos\phi_S(\cos\phi_S-\sqrt{2}\sin\phi_S)
\right.\nonumber\\
&& \left.
+\frac{m^2_K-m^2_{f_0}}{s-m^2_{f_0}}\sin\phi_S(\sin\phi_S+\sqrt{2}\cos\phi_S)\right]
\nonumber\\
&&
+\frac{1}{4f_\pi f_K}\left[(t-m^2_K)\frac{m^2_K-m^2_\kappa}{t-m^2_\kappa}+(t\leftrightarrow u)\right]\ .
\end{eqnarray}
One gets the $s$-dependent scalar amplitude in Eq.~(\ref{AKpKmpi0pi0ChPTLsM}) replacing the $t$- and $u$-dependent parts of the expression above by the corresponding chiral-loop amplitude
(\ref{AKpKmpi0pi0ChPT}) minus the infinite mass limit of the $s$-channel scalar contributions in
Eq.~(\ref{AKpKmpi0pi0LsMtu}).

\subsubsection*{\boldmath $K^+K^-\to\pi^0\eta$}
The $K^+K^-\to\pi^0\eta$ amplitude in the L$\sigma$M is
\begin{eqnarray}
{\cal A}_{K^+K^-\rightarrow\pi^0\eta}^{\mbox{\scriptsize L$\sigma$M}}
&=& g_{K^+ K^-\pi^0\eta}
         -\frac{g_{a_0 K^+K^-}g_{a_0\pi^0\eta}}{s-m^2_{a_{0}}}\nonumber\\
&&    -g_{\kappa^\mp K^\pm\pi^0}g_{\kappa^\mp K^\pm\eta}
         \left(\frac{1}{t-m^2_\kappa}+\frac{1}{u-m^2_\kappa}\right)\ ,
\end{eqnarray}
where the required $SPP$ couplings are
\begin{equation}
\label{gKpKmpi0eta}
g_{a_0\pi^0\eta}=\frac{m^2_\eta-m^2_{a_0}}{f_\pi}\cos\phi_P\ ,\quad
g_{\kappa^\mp K^\pm\eta}=\frac{m^2_\eta-m^2_\kappa}{2f_K}(\cos\phi_P-\sqrt{2}\sin\phi_P)\ ,
\end{equation}
and $\phi_P$ is the $\eta$-$\eta^\prime$ mixing angle in the quark-flavour basis defined as
\begin{equation}
\eta=\cos\phi_P\eta_q-\sin\phi_P\eta_s\ ,
\qquad
\eta^\prime=\sin\phi_P\eta_q+\cos\phi_P\eta_s\ ,
\end{equation}
with $\eta_q\equiv\frac{1}{\sqrt{2}}(u\bar u+d\bar d)$ and $\eta_s\equiv s\bar s$.
Using the soft-pion limit ($p_\pi\to 0$), which implies in this case $s=m^2_\eta$ and $t=u=m^2_K$,
the coupling $g_{K^+ K^-\pi^0\eta}$ is fixed to
\begin{equation}
g_{K^+ K^-\pi^0\eta}=\frac{g_{a_0 K^+K^-}g_{a_0\pi^0\eta}}{m^2_\eta-m^2_{a_{0}}}
+2\frac{g_{\kappa^\mp K^\pm\pi^0}g_{\kappa^\mp K^\pm\eta}}{m^2_K-m^2_\kappa}\ ,
\end{equation}
and the amplitude to
\begin{eqnarray}
{\cal A}_{K^+K^-\rightarrow\pi^0\eta}^{\mbox{\scriptsize L$\sigma$M}}
&=& \frac{g_{a_0 K^+K^-}g_{a_0\pi^0\eta}}{m^2_\eta-m^2_{a_{0}}}
         \frac{s-m^2_\eta}{s-m^2_{a_{0}}}\nonumber\\
&& +\frac{g_{\kappa^\mp K^\pm\pi^0}g_{\kappa^\mp K^\pm\eta}}{m^2_K-m^2_\kappa}
         \left(\frac{t-m^2_K}{t-m^2_\kappa}+ \frac{u-m^2_K}{u-m^2_\kappa}\right)\ .
\end{eqnarray}
Using the explicit expressions for the couplings in 
Eqs.~(\ref{gKpKmK0K0bar}), (\ref{gKpKmpi0pi0}) and (\ref{gKpKmpi0eta}),
the former amplitude is written as
\begin{eqnarray}
\label{AKpKmpi0etaLsMtu}
{\cal A}_{K^+K^-\rightarrow\pi^0\eta}^{\mbox{\scriptsize L$\sigma$M}}
&=& \frac{1}{2f_\pi f_K}\left\{(s-m^2_\eta)\frac{m^2_K-m^2_{a_0}}{s-m^2_{a_0}}\cos\phi_P
\right.\nonumber\\
&&
+\left[(t-m^2_K)\frac{m^2_\eta-m^2_\kappa}{t-m^2_\kappa}+
          (u-m^2_K)\frac{m^2_\eta-m^2_\kappa}{u-m^2_\kappa}\right]\nonumber\\
&&  \left.
\times\frac{1}{2}(\cos\phi_P-\sqrt{2}\sin\phi_P)\right\}\ .
\end{eqnarray}
Again, one gets the $s$-dependent scalar amplitude in Eq.~(\ref{AKpKmpi0etaChPTLsM})
replacing the $t$- and $u$-dependent parts of the expression above by the corresponding
chiral-loop amplitude (\ref{AKpKmpi0etaChPT}) minus the infinite mass limit of the
$s$-channel scalar contributions in Eq.~(\ref{AKpKmpi0etaLsMtu}).

\subsubsection*{\boldmath $\pi^+\pi^-\to\pi^0\pi^0$}
The $\pi^+\pi^-\to\pi^0\pi^0$ amplitude in the L$\sigma$M is
\begin{equation}
{\cal A}_{\pi^+\pi^-\rightarrow\pi^0\pi^0}^{\mbox{\scriptsize L$\sigma$M}}=
g_{\pi^+\pi^-\pi^0\pi^0}
-\frac{g_{\sigma\pi^+\pi^-}g_{\sigma\pi^0\pi^0}}{s-m^2_\sigma}
-\frac{g_{f_0\pi^+\pi^-}g_{f_0\pi^0\pi^0}}{s-m^2_{f_0}}\ ,
\end{equation}
where the required $SPP$ couplings are
\begin{equation}
\label{gpippimpi0pi0}
\begin{array}{c}
g_{\sigma\pi^+\pi^-}=g_{\sigma\pi^0\pi^0}=\frac{m^2_\pi-m^2_\sigma}{f_\pi}\cos\phi_S\ ,\\[2ex]
g_{f_0\pi^+\pi^-}=g_{f_0\pi^0\pi^0}=\frac{m^2_\pi-m^2_{f_0}}{f_\pi}\sin\phi_S\ ,
\end{array}
\end{equation}
and the soft-pion limit ($p_\pi\to 0$), which implies $s=t=u=m^2_\pi$,
fixes the coupling $g_{\pi^+\pi^-\pi^0\pi^0}$ to
\begin{equation}
g_{\pi^+\pi^-\pi^0\pi^0}=
\frac{g_{\sigma\pi^+\pi^-}g_{\sigma\pi^0\pi^0}}{m^2_\pi-m^2_\sigma}+
\frac{g_{f_0\pi^+\pi^-}g_{f_0\pi^0\pi^0}}{m^2_\pi-m^2_{f_0}}\ .
\end{equation}
Hence, the amplitude is written as
\begin{equation}
{\cal A}_{\pi^+\pi^-\rightarrow\pi^0\pi^0}^{\mbox{\scriptsize L$\sigma$M}}=
\frac{g_{\sigma\pi^+\pi^-}g_{\sigma\pi^0\pi^0}}{m^2_\pi-m^2_\sigma}
\frac{s-m^2_\pi}{s-m^2_\sigma}+
\frac{g_{f_0\pi^+\pi^-}g_{f_0\pi^0\pi^0}}{m^2_\pi-m^2_{f_0}}
\frac{s-m^2_\pi}{s-m^2_{f_0}}\ ,
\end{equation}
which turns out to be
\begin{equation}
\label{Apippimpi0pi0LsMs}
{\cal A}_{\pi^+\pi^-\rightarrow\pi^0\pi^0}^{\mbox{\scriptsize L$\sigma$M}}=
\frac{s-m^2_\pi}{f_\pi^2}
\left(\frac{m^2_\pi-m^2_\sigma}{s-m^2_\sigma}\cos^2\phi_S+
        \frac{m^2_\pi-m^2_{f_0}}{s-m^2_{f_0}}\sin^2\phi_S\right)\ ,
\end{equation}
once the explicit expressions in Eq.~(\ref{gpippimpi0pi0}) are used.
As seen, this amplitude is already $s$-dependent and coincides with its chiral-loop counterpart
(\ref{Apippimpi0pi0ChPT}) in the infinite scalar mass limit.
Hence, it can be directly plugged into
${\cal A}_{\rho\rightarrow\pi^0\pi^0\gamma}^{\mbox{\scriptsize L$\sigma$M}}$
to get the total scalar contribution to this process.

\subsection{Complete one-loop propagators}
\label{propagators}
The complete one-loop propagators for the $f_0$ and $a_0$ scalar resonances are defined as
\begin{equation}
\label{propagator}
D(s)=s-m_R^2+{\rm Re}\Pi(s)-{\rm Re}\Pi(m_R^2)+ i{\rm Im}\Pi(s)\ ,
\end{equation}
where $m_R$ is the renormalized mass of the scalar meson and
$\Pi(s)$ is the one-particle irreducible two-point function.
${\rm Re}\Pi(m_R^2)$ is introduced to regularize the divergent behaviour of $\Pi(s)$.
The propagator so defined is well behaved when a threshold is approached from below,
thus improving the usual Breit-Wigner prescription not particularly suited for spinless resonances
(see Ref.~\cite{Escribano:2002iv} for details).
However, for the $\sigma$ meson we choose a simple Breit-Wigner propagator
since the values of mass and width used in our analysis are obtained from experimental data
applying this parametrization.
The real and imaginary parts of $\Pi(s)$ for the $f_0$ are\footnote{
In our analysis, we work in the isospin limit and therefore the mass difference
between $K^0$ and $K^+$ is not taken into account for the $K\bar K$ threshold.}
$(R(s)\equiv {\rm Re}\Pi(s),I(s)\equiv {\rm Im}\Pi(s))$
\begin{equation}
\label{RI}
\begin{array}{l}
\nonumber
R(s)=\frac{g_{f_0\pi\pi}^2}{16\pi^2}
\left[2-\beta_\pi\log\left(\frac{1+\beta_\pi}{1-\beta_\pi}\right)\Theta_\pi
-2\bar\beta_\pi\arctan\left(\frac{1}{\bar\beta_\pi}\right)\bar\Theta_\pi\right]\\[2ex]
\qquad\ 
+\frac{g_{f_0 K\bar K}^2}{16\pi^2}
\left[2-\beta_K\log\left(\frac{1+\beta_K}{1-\beta_K}\right)\Theta_K
-2\bar\beta_K\arctan\left(\frac{1}{\bar\beta_K}\right)\bar\Theta_K\right]\ ,\\[2ex]
I(s)=\frac{g_{f_0\pi\pi}^2}{16\pi}\beta_\pi\Theta_\pi+\frac{g_{f_0 K\bar K}^2}{16\pi}\beta_K\Theta_K\ ,
\end{array}
\end{equation}
where $\beta_i=\sqrt{1-4m_i^2/s}$ for $i=(\pi, K)$,
$\bar\beta_i=\sqrt{4m_i^2/s-1}$, $\Theta_i=\Theta(s-4m_i^2)$, and 
$\bar\Theta_i=\Theta(4m_i^2-s)$.
The couplings of the $f_0$ to pions and kaons are written in the isospin limit so
$g_{f_0\pi\pi}^2=\frac{3}{2}g_{f_0\pi^+\pi^-}^2$ and $g_{f_0 K\bar K}^2=2g_{f_0 K^+K^-}^2$.
For the $a_0$ the real and imaginary parts of the two-point function are
\begin{equation}
\label{RI2}
\begin{array}{l}
\nonumber
R(s)=\frac{g_{f_0 K\bar K}^2}{16\pi^2}
\left[2-\beta_K\log\left(\frac{1+\beta_K}{1-\beta_K}\right)\Theta_K
-2\bar\beta_K\arctan\left(\frac{1}{\bar\beta_K}\right)\bar\Theta_K\right]\\[2ex]
\qquad\
+\frac{g_{a_0\pi\eta}^2}{16\pi^2}
\left[2-\frac{m^2_\eta-m^2_\pi}{s}\log\left(\frac{m_\eta}{m_\pi}\right)
-\beta^+_{\pi\eta}\beta^-_{\pi\eta}
\log\left(\frac{\beta^-_{\pi\eta}+\beta^+_{\pi\eta}}{\beta^-_{\pi\eta}-\beta^+_{\pi\eta}}\right)
\Theta_{\pi\eta}\right.\\[2ex]
\qquad\quad\left.
-2\bar\beta^+_{\pi\eta}\beta^-_{\pi\eta}\arctan\left(\frac{\beta^-_{\pi\eta}}{\bar\beta^+_{\pi\eta}}\right)
\bar\Theta_{\pi\eta}
+\bar\beta^+_{\pi\eta}\bar\beta^-_{\pi\eta}
\log\left(\frac{\bar\beta^-_{\pi\eta}+\bar\beta^+_{\pi\eta}}{\bar\beta^-_{\pi\eta}-\bar\beta^+_{\pi\eta}}\right)
\bar{\bar\Theta}_{\pi\eta}\right]\ ,\\[2ex]
I(s)=\frac{g_{a_0 K\bar K}^2}{16\pi}\beta_K\Theta_K
      +\frac{g_{a_0\pi\eta}^2}{16\pi}\beta^+_{\pi\eta}\beta^-_{\pi\eta}\Theta_{\pi\eta}\ ,
\end{array}
\end{equation}
where $\beta^\pm_{\pi\eta}=\sqrt{1-(m_\pi\pm m_\eta)^2/s}$,
$\bar\beta^\pm_{\pi\eta}=\sqrt{(m_\pi\pm m_\eta)^2/s-1}$,
$\Theta_{\pi\eta}=\Theta[s-(m_\pi+m_\eta)^2]$,
$\bar\Theta_{\pi\eta}=\Theta[s-(m_\pi-m_\eta)^2]\times\Theta[(m_\pi+m_\eta)^2-s]$, and 
$\bar{\bar\Theta}_{\pi\eta}=\Theta[(m_\pi-m_\eta)^2-s]$.

\subsection{Invariant mass distribution}
\label{invmassdis}
The final $P^0P^0$ invariant mass distribution for the $V\to P^0P^0\gamma$ processes is\footnote{
In terms of the photon energy, $E_\gamma=(m^2_{V^0}-m^2_{P^0P^{0\prime}})/(2m_{V^0})$,
the photonic spectrum is written as
$d\Gamma/dE_\gamma=(m_{V^0}/m_{P^0P^{0\prime}})\times d\Gamma/dm_{P^0P^{0\prime}}$.}
\begin{equation}
\label{dGdmtotal}
\frac{d\Gamma(V\rightarrow P^0P^0\gamma)}{dm_{P^0P^{0\prime}}}=
\frac{d\Gamma_{\mbox{\scriptsize L$\sigma$M}}}{dm_{P^0P^{0\prime}}}+
\frac{d\Gamma_{\mbox{\scriptsize VMD}}}{dm_{P^0P^{0\prime}}}+
\frac{d\Gamma_{\mbox{\scriptsize int}}}{dm_{P^0P^{0\prime}}}\ ,
\end{equation}
where the L$\sigma$M term (the scalar or \emph{scalar} contribution) is\footnote{
For the generic process $V\to P^0P^0\gamma$ the notation is fixed to define
$m_{P^0}$ as the mass of the first pseudoscalar in the reaction and
$m_{P^{0\prime}}$ the mass of the second one.}
\begin{equation}
\label{dGLsigmaM}
\begin{array}{rcl}
\frac{d\Gamma_{\mbox{\tiny L$\sigma$M}}}{dm_{P^0P^{0\prime}}}&=&
\left(\frac{1}{2},1\right)\frac{\alpha}{192\pi^5}\frac{(g,g_{s})^2}{4\pi}\frac{m^4_{V^0}}{m^4_{P^+}}
\frac{m_{P^0P^{0\prime}}}{m_{V^0}}\left(1-\frac{m^2_{P^0P^{0\prime}}}{m^2_{V^0}}\right)^3]\\[2ex]
&\times &
\sqrt{\left(1-\frac{(m_{P^0}+m_{P^{0\prime}})^2}{m^2_{P^0P^{0\prime}}}\right)
         \left(1-\frac{(m_{P^0}-m_{P^{0\prime}})^2}{m^2_{P^0P^{0\prime}}}\right)}\\[2ex]
&\times & C^2_{\mbox{\scriptsize L$\sigma$M}} |L(m^2_{P^0P^{0\prime}})|^2
|{\cal A}(P^+P^-\rightarrow P^0P^0)_{\mbox{\scriptsize L$\sigma$M}}|^2\ ,
\end{array}
\end{equation}
with $\left(\frac{1}{2},1\right)$ for the case or not of identical pseudoscalar states in the final state,
$(g,g_{s})$ for the case or not of the $\phi$ meson as the initial decaying vector,
$m_P^+$ the mass of the charged pseudoscalar inside the loop
($m_{\pi^+}$ for $\rho\to\pi^0\pi^0\gamma$, $m_{K^+}$ otherwise),
$C_{\mbox{\scriptsize L$\sigma$M}}$ the coefficients defined as
\begin{equation}
\label{CLsM}
\begin{array}{c}
C_{\mbox{\scriptsize L$\sigma$M}}^{\phi\to\pi^0\pi^0\gamma}=
-\sqrt{2}C_{\mbox{\scriptsize L$\sigma$M}}^{\omega\to\pi^0\pi^0\gamma}=
-\frac{1}{\sqrt{2}}C_{\mbox{\scriptsize L$\sigma$M}}^{\rho\to\pi^0\pi^0\gamma}=1\ ,\\[1ex]
C_{\mbox{\scriptsize L$\sigma$M}}^{\phi\to\pi^0\eta\gamma}=
-\sqrt{2}C_{\mbox{\scriptsize L$\sigma$M}}^{\omega\to\pi^0\eta\gamma}=
-\sqrt{2}C_{\mbox{\scriptsize L$\sigma$M}}^{\rho\to\pi^0\eta\gamma}=1\ ,\\[1ex]
C_{\mbox{\scriptsize L$\sigma$M}}^{\phi\to K^0\bar K^0\gamma}=1\ ,
\end{array}
\end{equation}
and
${\cal A}(P^+P^-\rightarrow P^0P^0)_{\mbox{\scriptsize L$\sigma$M}}$
the four-pseudoscalar amplitudes taken from Eq.~(\ref{AKpKmK0K0barChPTLsM})
and Eqs.~(\ref{AKpKmpi0pi0ChPTLsM}--\ref{Apippimpi0pi0ChPTLsM}).
The VMD term (the vector or \emph{background} contribution)
and the contribution resulting from the interference of both amplitudes are given by
\begin{equation}
\label{dGdmVMDint}
\begin{array}{rcl}
\frac{d\Gamma_{\mbox{\tiny [VMD,int]}}}{dm_{P^0P^{0\prime}}}&=&
\left(\frac{1}{2},1\right)\frac{1}{256\pi^3}
\frac{m_{P^0P^{0\prime}}}{m_{V^0}}\left(1-\frac{m^2_{P^0P^{0\prime}}}{m^2_{V^0}}\right)\\[2ex]
&\times &
\sqrt{\left(1-\frac{(m_{P^0}+m_{P^{0\prime}})^2}{m^2_{P^0P^{0\prime}}}\right)
         \left(1-\frac{(m_{P^0}-m_{P^{0\prime}})^2}{m^2_{P^0P^{0\prime}}}\right)}\\[2ex]
&\times &
{\displaystyle\int_{-1}^{1}}dx\,A_{\rm [VMD,int]}(m_{P^0P^{0\prime}},x)\ ,
\end{array}
\end{equation}
where one explicitly has
\begin{equation}
\label{AVMD2}
\begin{array}{l}
A_{\rm VMD}(m^2_{P^0P^{0\prime}},x)\equiv 
\frac{1}{3}\sum_{\rm pol}|{\cal A}_{\rm VMD}|^2=
\frac{1}{3}\left(C_{\rm VMD}\frac{G^2 e}{\sqrt{2}g}\right)^2\\[2ex]
\quad\times
\left\{\frac{1}{8}\left[
m^4_{P^0}m^4_{P^{0\prime}}-2m^2_{V^0}m^4_{P^0}m^2_{P^{0\prime}}
+(m^2_{V^0}-\tilde m^2_V)^2m^4_{P^0}+\tilde m^4_Vm^4_{P^{0\prime}}\right.\right.\\[2ex]
\qquad
+2\tilde m^2_V(m^2_{V^0}+\tilde m^2_V-\tilde m^{\ast 2}_{V^\prime})m^2_{P^0}m^2_{P^{0\prime}}
-2\tilde m^4_V(\tilde m^2_V+\tilde m^{\ast 2}_{V^\prime})m^2_{P^{0\prime}}\\[2ex]
\qquad
-2\tilde m^2_V(m^2_{V^0}-\tilde m^2_V)(m^2_{V^0}-\tilde m^2_V-\tilde m^{\ast 2}_{V^\prime})m^2_{P^0}
+\tilde m^4_V\tilde m^{\ast 4}_{V^\prime}\\[2ex]
\qquad\left.
+\tilde m^4_V(m^2_{V^0}+\tilde m^2_V-\tilde m^{\ast 2}_{V^\prime})^2\right]
\times\frac{1}{|D_{V}(\tilde m^2_V)|^2}\\[2ex]
\quad
+\frac{1}{16}\left[
m^4_{P^0}m^4_{P^{0\prime}}-\tilde m^{\ast 2}_{V^\prime}(m^2_{V^0}-\tilde m^2_V)^2m^4_{P^0}
-\tilde m^2_V(m^2_{V^0}-\tilde m^{\ast 2}_{V^\prime})^2m^4_{P^{0\prime}}\right.\\[2ex]
\qquad
+2\tilde m^2_V\tilde m^{\ast 2}_{V^\prime}(m^2_{V^0}-\tilde m^2_V-\tilde m^{\ast 2}_{V^\prime})
(m^2_{P^0}+m^2_{P^{0\prime}})+\tilde m^4_V\tilde m^{\ast 4}_{V^\prime}\\[2ex]
\qquad\left.
-\tilde m^2_V\tilde m^{\ast 2}_{V^\prime}(\tilde m^2_V+\tilde m^{\ast 2}_{V^\prime})
(m^2_{V^0}-\tilde m^2_V-\tilde m^{\ast 2}_{V^\prime})\right]\\[2ex]
\qquad\left.
\times
2\mbox{Re}\left(\frac{1}{D_{V}(\tilde m^2_V)D^\ast_{V^\prime}(\tilde m^{\ast 2}_{V^\prime})}\right)
+\left(m^2_{P^0},\tilde m^2_V,V\right)\leftrightarrow
  \left(m^2_{P^{0\prime}},\tilde m^{\ast 2}_{V^\prime},V^\prime\right)\right\}\ ,
\end{array}
\end{equation}
and 
\begin{equation}
\label{Aint}
\begin{array}{l}
A_{\rm int}(m^2_{P^0P^{0\prime}},x)\equiv
\frac{2}{3}\mbox{Re}\sum_{\rm pol}{\cal A}_{\mbox{\scriptsize L$\sigma$M}}{\cal A}_{\rm VMD}^\ast\\[2ex]
\qquad
=\frac{1}{3}\left(C_{\mbox{\scriptsize L$\sigma$M}}\frac{e(g,g_{s})}{2\pi^2 m^2_{P^+}}\right)
\left(C_{\rm VMD}\frac{G^2 e}{\sqrt{2}g}\right)\\[2ex]
\qquad
\times 2\mbox{Re}\left\{L(m^2_{P^0P^{0\prime}})
{\cal A}(P^+P^-\rightarrow P^0P^{0\prime})_{\mbox{\scriptsize L$\sigma$M}}\right.\\[2ex]
\qquad
\times\frac{1}{4}\left[
\frac{\tilde m^2_V(\tilde m^2_V+\tilde m^{\ast 2}_{V^\prime}-m^2_{P^0}-m^2_{P^{0\prime}})^2
         -m^2_{V^0}(\tilde m^2_V-m^2_{P^0})^2}{D^\ast_{V}(\tilde m^2_V)}\right.\\[2ex]
\qquad\left.\left.
+\left(m^2_{P^0},\tilde m^2_V,V\right)\leftrightarrow
  \left(m^2_{P^{0\prime}},\tilde m^{\ast 2}_{V^\prime},V^\prime\right)\right]\right\}\ ,
\end{array}
\end{equation}
with $(P^2\equiv\tilde m^2_V,{P^\prime}^2\equiv\tilde m^{\ast 2}_{V^\prime})$
\begin{equation}
\label{mVtilde}
\begin{array}{l}
\tilde m^2_V=
m^2_{P^0}+\frac{m^2_{V^0}-m^2_{P^0P^{0\prime}}}{2}
\left[1+\frac{m^2_{P^0}-m^2_{P^{0\prime}}}{m^2_{P^0P^{0\prime}}}
       -\left(1+\frac{m^2_{P^0}-m^2_{P^{0\prime}}}{m^2_{P^0P^{0\prime}}}\right)\beta_{P^0}x\right]\ ,\\[2ex]
\tilde m^{\ast 2}_{V^\prime}=
m^2_{P^{0\prime}}+\frac{m^2_{V^0}-m^2_{P^0P^{0\prime}}}{2}
\left[1-\frac{m^2_{P^0}-m^2_{P^{0\prime}}}{m^2_{P^0P^{0\prime}}}
       +\left(1+\frac{m^2_{P^0}-m^2_{P^{0\prime}}}{m^2_{P^0P^{0\prime}}}\right)\beta_{P^0}x\right]\ ,\\[2ex]
\qquad
=m^2_{P^0}+m^2_{P^{0\prime}}+m^2_{V^0}-m^2_{P^0P^{0\prime}}-\tilde m^2_V\ ,
\end{array}
\end{equation}
and
\begin{equation}
\label{betaP0}
\beta_{P^0}=\sqrt{1-\frac{4m^2_{P^0}}{m^2_{P^0P^{0\prime}}}
                        \frac{1}{\left(1+\frac{m^2_{P^0}-m^2_{P^{0\prime}}}{m^2_{P^0P^{0\prime}}}\right)^2}}\ .
\end{equation}

The invariant mass distribution for the processes with $m^2_{P^0}=m^2_{P^{0\prime}}$
can be simplified to
\begin{equation}
\label{dGLsigmaMmeq}
\begin{array}{rcl}
\frac{d\Gamma_{\mbox{\tiny L$\sigma$M}}}{dm_{P^0P^0}}&=&
\frac{1}{2}\frac{\alpha}{192\pi^5}\frac{(g,g_{s})^2}{4\pi}\frac{m^4_{V^0}}{m^4_{P^+}}
\frac{m_{P^0P^0}}{m_{V^0}}\left(1-\frac{m^2_{P^0P^0}}{m^2_{V^0}}\right)^3
\sqrt{1-\frac{4m^2_{P^0}}{m^2_{P^0P^0}}}\\[2ex]
&\times & C^2_{\mbox{\scriptsize L$\sigma$M}} |L(m^2_{P^0P^0})|^2
|{\cal A}(P^+P^-\rightarrow P^0P^0)_{\mbox{\scriptsize L$\sigma$M}}|^2\ ,
\end{array}
\end{equation}
for the scalar contribution, and
\begin{equation}
\label{dGdmVMDintmeq}
\begin{array}{rcl}
\frac{d\Gamma_{\mbox{\tiny [VMD,int]}}}{dm_{P^0P^0}}&=&
\frac{1}{2}\frac{1}{256\pi^3}
\frac{m_{P^0P^0}}{m_{V^0}}\left(1-\frac{m^2_{P^0P^0}}{m^2_{V^0}}\right)^3
\sqrt{1-\frac{4m^2_{P^0}}{m^2_{P^0P^0}}}\\[2ex]
&\times &
{\displaystyle\int_{-1}^{1}}dx\,A_{\rm [VMD,int]}(m_{P^0P^0},x)\ ,
\end{array}
\end{equation}
where
\begin{equation}
\label{AVMD2meq} 
\begin{array}{l}
A_{\rm VMD}(m^2_{P^0P^0},x)\equiv 
\frac{1}{3}\sum_{\rm pol}|{\cal A}_{\rm VMD}|^2=
\frac{1}{3}\left(C_{\rm VMD}\frac{G^2 e}{\sqrt{2}g}\right)^2\\[2ex]
\quad\times
\left\{\frac{1}{8}\left[
m^8_{P^0}-2m^2_{V^0}m^6_{P^0}
+(m^4_{V^0}+4\tilde m^4_V-2\tilde m^2_V\tilde m^{\ast 2}_{V^\prime})m^4_{P^0}\right.\right.\\[2ex]
\qquad
-2\tilde m^2_Vm^2_{V^0}(m^2_{V^0}-2\tilde m^2_V-\tilde m^{\ast 2}_{V^\prime})m^2_{P^0}
-4\tilde m^4_V(\tilde m^2_V+\tilde m^{\ast 2}_{V^\prime})m^2_{P^0}\\[2ex]
\qquad\left.
+\tilde m^4_V\tilde m^{\ast 4}_{V^\prime}
+\tilde m^4_V(m^2_{V^0}+\tilde m^2_V-\tilde m^{\ast 2}_{V^\prime})^2\right]
\times\frac{1}{|D_{V}(\tilde m^2_V)|^2}\\[2ex]
\quad
+\frac{1}{16}\left[
m^8_{P^0}+2\tilde m^2_V\tilde m^{\ast 2}_{V^\prime}m^4_{P^0}
-m^2_{V^0}(\tilde m^2_V+\tilde m^{\ast 2}_{V^\prime})m^4_{P^0}\right.\\[2ex]
\qquad
+4\tilde m^2_V\tilde m^{\ast 2}_{V^\prime}
(m^2_{V^0}-\tilde m^2_V-\tilde m^{\ast 2}_{V^\prime})m^2_{P^0}
+\tilde m^4_V\tilde m^{\ast 4}_{V^\prime}\\[2ex]
\qquad\left.
-\tilde m^2_V\tilde m^{\ast 2}_{V^\prime}(\tilde m^2_V+\tilde m^{\ast 2}_{V^\prime})
(m^2_{V^0}-\tilde m^2_V-\tilde m^{\ast 2}_{V^\prime})\right]\\[2ex]
\qquad\left.
\times
2\mbox{Re}\left(\frac{1}{D_{V}(\tilde m^2_V)D^\ast_{V^\prime}(\tilde m^{\ast 2}_{V^\prime})}\right)
+\left(\tilde m^2_V,V\right)\leftrightarrow
  \left(\tilde m^{\ast 2}_{V^\prime},V^\prime\right)\right\}\ ,
\end{array}
\end{equation}
and 
\begin{equation}
\label{Aintmeq}
\begin{array}{l}
A_{\rm int}(m^2_{P^0P^0},x)\equiv
\frac{2}{3}\mbox{Re}\sum_{\rm pol}{\cal A}_{\mbox{\scriptsize L$\sigma$M}}{\cal A}_{\rm VMD}^\ast\\[2ex]
\qquad
=\frac{1}{3}\left(C_{\mbox{\scriptsize L$\sigma$M}}\frac{e(g,g_{s})}{2\pi^2 m^2_{P^+}}\right)
\left(C_{\rm VMD}\frac{G^2 e}{\sqrt{2}g}\right)\\[2ex]
\qquad
\times 2\mbox{Re}\left\{L(m^2_{P^0P^0})
{\cal A}(P^+P^-\rightarrow P^0P^0)_{\mbox{\scriptsize L$\sigma$M}}\right.\\[2ex]
\qquad
\times\frac{1}{4}\left[
\frac{\tilde m^2_V(\tilde m^2_V+\tilde m^{\ast 2}_{V^\prime}-2m^2_{P^0})^2
         -m^2_{V^0}(\tilde m^2_V-m^2_{P^0})^2}{D^\ast_{V}(\tilde m^2_V)}\right.\\[2ex]
\qquad\left.\left.
+\left(\tilde m^2_V,V\right)\leftrightarrow
  \left(\tilde m^{\ast 2}_{V^\prime},V^\prime\right)\right]\right\}\ ,
\end{array}
\end{equation}
with
\begin{equation}
\label{mVtildemeq}
\begin{array}{l}
\tilde m^2_V=
m^2_{P^0}+\frac{m^2_{V^0}-m^2_{P^0P^0}}{2}
\left(1-x\sqrt{1-\frac{4m^2_{P^0}}{m^2_{P^0P^0}}}\right)\ ,\\[2ex]
\tilde m^{\ast 2}_{V^\prime}=
m^2_{P^0}+\frac{m^2_{V^0}-m^2_{P^0P^0}}{2}
\left(1+x\sqrt{1-\frac{4m^2_{P^0}}{m^2_{P^0P^0}}}\right)\ ,\\[2ex]
\qquad
=2m^2_{P^0}+m^2_{V^0}-m^2_{P^0P^0}-\tilde m^2_V\ .\\[2ex]
\end{array}
\end{equation}

\end{document}